\documentclass[lettersize,journal]{IEEEtran}
\usepackage{hyperref}
\usepackage[english]{babel}
\usepackage{amsmath}
\usepackage{amssymb}
\usepackage{graphicx}
\usepackage{wrapfig}
\usepackage{verbatim}
\let\labelindent\relax
\usepackage{enumitem}
\usepackage{threeparttable}
\usepackage{multirow}
\usepackage{xspace}
\usepackage{xcolor}
\usepackage{xurl}
\usepackage{tabularx}
\usepackage{soul}
\usepackage{pgfkeys,pgfmath}  
\usepackage{siunitx}
\usepackage{subcaption}
\usepackage[ruled,linesnumbered]{algorithm2e}
\usepackage{cite}
\usepackage{authblk}
\usepackage[table]{xcolor} % for row coloring
\usepackage{booktabs}      % for nice horizontal rules
\usepackage{colortbl}      % enables \rowcolor

\RestyleAlgo{ruled}
\SetKwInput{KwData}{Input} 
\SetKwInput{KwResult}{Output}

\usepackage[colorinlistoftodos]{todonotes}
\usepackage{booktabs}
\setitemize{noitemsep,topsep=0pt,parsep=0pt,partopsep=0pt}
\renewcommand\IEEEkeywordsname{Index Terms}

\SetKwComment{Comment}{> }{}

\newcommand{\rev}[1]{#1}
\newcommand{\mypar}[1]{\noindent\textbf{#1}}
\newcommand{\vct}[1]{\ensuremath{\mathbf{#1}}}

\newcommand{\smalldata}{\ensuremath{\mathcal{D}_{s}}\xspace}
\newcommand{\largedata}{\ensuremath{\mathcal{D}_{l}}\xspace}
\newcommand{\synthVTCardinality}{160\xspace}
\newcommand{\goodwareCardinality}{20,769\xspace}
\newcommand{\malwareCardinality}{20,329\xspace}

\newcommand{\repackagedCardinality}{20,284\xspace}
\begin{document}

% \date{}

%\pagestyle{empty}
% \title{Manipulating Trust: The Threat of Label Spoofing Attacks in Machine Learning-Based Malware Classifiers\\
% \textcolor{blue}{\\
% or\\
% Stealth Malware: Exploiting Label Spoofing to Poison Machine Learning Models}
% }

\title{Trust Under Siege: Label Spoofing Attacks Against Machine Learning for Android Malware Detection}

\author[1]{Tianwei Lan}
\author[2]{Luca Demetrio}
\author[1]{Farid Nait-Abdesselam}
\author[3]{Yufei Han}
\author[4]{Simone Aonzo}

\affil[1]{Université Paris Cité, France}
\affil[2]{University of Genova, Italy}
\affil[3]{INRIA, France}
\affil[4]{EURECOM, France}

\maketitle
\begin{abstract}
Machine Learning (ML) malware detectors rely heavily on crowd-sourced AntiVirus (AV) labels, with platforms like VirusTotal serving as trusted sources of malware annotations. 
But what if attackers could manipulate these labels to classify benign software as malicious?
We introduce label spoofing attacks, a new threat that contaminates crowd-sourced datasets by embedding minimal and undetectable malicious patterns into benign samples. These patterns coerce AV engines into misclassifying legitimate files as harmful, enabling poisoning attacks against ML-based malware classifiers trained on those data.
We demonstrate this scenario by developing AndroVenom, a methodology for polluting realistic data sources and launching subsequent poisoning attacks against ML malware detectors. 
Experiments show that not only are state-of-the-art feature extractors unable to filter such injections, but various ML models experience Denial-of-Service (DoS) with as little as 1\% poisoned samples.
Additionally, attackers can flip decisions for specific unaltered benign samples by modifying only 0.015\% of the training data, threatening their reputation and market share, while evading anomaly detectors operating on the training data.
We conclude by raising concerns about the trustworthiness of ML training processes based on AV annotations and argue that further investigation is needed to develop more reliable labeling strategies.
\end{abstract}


\pagestyle{plain}
\section{Introduction}

\IEEEPARstart{I}{n} response to the proliferation of malware samples, both the security research community and industrial security service providers increasingly rely on Machine Learning (ML) techniques to discriminate malicious from benign applications~\cite{avtest_mstatistics}. 
It is achieved by capturing correlations between features extracted from samples and their corresponding labels within the training dataset. 
Hence, the quality of ML-based malware detectors directly depends on the availability of reliable ground truth for both benign and malicious applications in the training set.
Over time, the community has refined best practices for constructing high-quality datasets for training purposes.
To collect benign applications, researchers and practitioners often acquire samples from well-monitored repositories. 
In the case of Android applications (APKs), benign samples are typically fetched from the Google Play Store~\cite{ruggia24unmasking}, where each application undergoes rigorous vetting processes that minimize the presence of malware on the platform.
Conversely, for Windows programs, data is often extracted from fresh installations of community-maintained packages hosted on Chocolatey, which also enforces strict moderation and review procedures to prevent contamination~\cite{dambra2023decoding}.
To collect malicious APKs, security service providers and practitioners turn to online crowdsourcing platforms such as VirusTotal~\cite{virustotal,zhu2020measuring}, which aggregates scan results from over 70 commercial AntiVirus (AV) engines~\cite{vt_contributors} while storing each submitted sample. 
Samples are typically included in datasets as malicious when the number of AV detections exceeds a predefined threshold~\cite{ruggia2022,ruggia24unmasking,ArpSHGR14,dambra2023decoding,aonzo2023humans,zhu2020measuring}.
However, while the use of these platforms streamlines and automates malware annotation, it also introduces a subtle yet critical vulnerability.
Commercial AVs often rely on signature-based techniques~\cite{koret2015antivirus}, detecting malware by identifying specific artifacts such as cryptographic hashes, unique strings, or characteristic code patterns.
This reliance enables an attacker to strategically embed recognized \emph{malicious byte patterns} into otherwise benign programs, forcing AV engines to flag them as malicious.
Notably, injected signatures, such as the EICAR signature~\cite{eicar}, do not introduce new malicious functionality, and the modified programs do not exhibit harmful behavior at runtime.
Once uploaded to VirusTotal or similar antivirus detection services, these modified samples are labeled as malicious solely due to the presence of the injected signature.
This situation creates a fundamental ambiguity: while AVs correctly identify a known malicious artifact, the program’s functionality remains benign.
We define and exploit this ambiguity as a \emph{label spoofing attack}.
When downstream users retrieve such mislabeled samples from crowdsourcing platforms and incorporate them into ML training datasets, attackers gain indirect influence over the resulting models.
Specifically, they can stage \emph{poisoning attacks}~\cite{biggio2018wild, cina2023wild}, ranked among the most impactful threats to ML systems by OWASP~\cite{owasp_topten}. 
These attacks can either render malware detectors unusable (\emph{Denial of Service}) or trigger excessive false alarms against targeted benign applications, damaging their reputation and hindering their distribution.
Such degradation can lead to \emph{(i)} erosion of trust in crowdsourced malware annotations, \emph{(ii)} financial losses for AV vendors due to increased false positives, and \emph{(iii)} reputational harm to software developers whose products are repeatedly misclassified as malicious.
In extreme scenarios, this mechanism could be abused by malicious actors or state entities to censor legitimate software under the guise of national security by deliberately polluting training datasets.
Furthermore, human analysts may waste significant time investigating spoofed samples that contain no real malicious behavior.
To analyze the threat posed by label spoofing attacks against ML-based malware classifiers, we investigate the following research questions:

\mypar{RQ1.} How can an attacker inject a malicious signature with \emph{minimal} manipulation that remains undetected by feature extractors?

\mypar{RQ2.} What conditions enable label spoofing attacks, and what level of effort is required for an attacker to tamper with the labeling systems of AVs?

\mypar{RQ3.} How many poisoned benign samples are required to launch Denial of Service (DoS) attacks that degrade the performance of ML-based malware detectors in production?

\mypar{RQ4.} Can label spoofing attacks induce targeted false alarms on selected benign samples while preserving overall malware classification accuracy?

We answer these questions by exposing the risks of over-reliance on AV labeling through the development of \emph{AndroVenom}, a framework that exploits this weakness and paves the way for poisoning attacks against ML malware detectors (\autoref{sec:androvenom}). 
Focusing on Android, given its unrivaled market share~\cite{osshare}, we address \textbf{RQ1} by introducing a novel manipulation of Android Applications (APKs). 
This manipulation is negligible in size relative to the original application yet causes legitimate samples to be flagged as malicious by an average of 20 AV engines, a threshold commonly considered indicative of true malware~\cite{zhu2020measuring}.
We further demonstrate that this modification remains \emph{invisible} to state-of-the-art feature extraction pipelines, including Drebin, MaMaDroid, and MalScan~\cite{ArpSHGR14,onwuzurike2019mamadroid, MalScan2020}, as their preprocessing outputs remain unchanged.
We then address \textbf{RQ2} by demonstrating the practical feasibility of label spoofing attacks on Android APKs (\autoref{subsec:feasability}). 
Repackaged applications are consistently flagged as malicious because AV engines identify the injected artifact via its cryptographic hash and associate it with attacker-selected malware families.
Finally, we answer \textbf{RQ3} and \textbf{RQ4} through extensive experimental evaluation (\autoref{sec:untargeted} and \autoref{sec:targeted}).
Using a dataset of 40K samples to emulate a realistic ML malware classification pipeline using Drebin, MaMaDroid, and MalScan feature extractors, we demonstrate that poisoning just 1\% of the training set can degrade detector performance by 15\% in low-false-positive regimes.
More alarmingly, only 0.015\% poisoned samples are sufficient to induce targeted misclassifications of specific benign applications while maintaining overall performance.
Lastly, we show that removing label-spoofed samples is non-trivial: ad-hoc defenses against clean-label poisoning can introduce performance degradation unacceptable in production environments.
While poisoning attacks are already well studied, AndroVenom highlights the overlooked risks stemming from blind trust in crowdsourced AV annotations—an issue that demands joint attention from both ML practitioners and AV vendors.
 
\section{Background}\label{sec:background}
Before delving into the details of our work, we summarize the core concepts needed to understand our methodology. 

\mypar{Android Applications.} Mobile apps running on Android are distributed in the form of App Package (APK) files, which are an extension of the Java JAR format.
An APK contains several files and folders, and in particular: 
\emph{(i)} \texttt{AndroidManifest.xml}, a file needed to identify and run the APK, containing metadata, such as its name, version, permissions, and high-level components; 
\emph{(ii)} \texttt{classes.dex}, files that contain the compiled bytecode, usually written in Java or Kotlin;
\emph{(iii)} \texttt{lib}, a folder containing the native libraries of the APK written in C or C++, compiled for different architectures (ARM and x86), and loaded through the Java Native Interface (JNI);
\emph{(iv)} \texttt{assets}, a folder containing resources that do not require compilation, such as texts, XML, fonts, music, and videos;
\emph{(v)} \texttt{res}, a folder of resources that require compilation, like XML layouts, images, strings, and colors.

\mypar{AntiVirus Software (AV).} This family of software is tasked with detecting (binary classification) and identifying (family classification) malicious activity from an input byte stream (e.g., a single file or network traffic).
It plays a vital role in defending modern IT systems ~\cite{avcomp,avtest_mstatistics,clamav,min2014antivirus,koret2015antivirus,zhu2020measuring}.
Being composed of different layers of defenses, one key component of AVs is the collection of \emph{signatures}, used to spot already-known threats.
Signatures are typically a collection of cryptographic hashes or byte streams extracted from previously collected malware samples~\cite{yara}.
It is usually the responsibility of human analysts to write these signatures when they encounter new malicious code, but they can also be generated automatically~\cite{naik2020evaluating}. 
With efficient pattern-matching algorithms, signatures are tested on input samples, and those are flagged as malicious once matches are found. 
Due to their speed and efficiency, detection through pattern matching is the most typical and preferred method for AVs. 
\rev{Crucially, their outputs are frequently reused as labels for training ML-based malware detectors, creating a transitive trust dependency between AV decisions and ML training data.}

\mypar{APK Repackaging.} It refers to the practice of extracting contents from an APK, modifying it, and recreating another valid APK. 
There are several contexts in which repackaging can be used, including reverse engineering tasks or illegally cloning APKs from the Play Store.
While in theory this operation is always possible,  Apktool~\cite{apktool}, the primary tool used for this operation, is obviously not free of bugs. 

\mypar{Android Malware Classification with ML.}
There are multiple ways to detect Android malware with ML techniques.
In this study, we focus on \emph{static analysis}, which relies on features extracted from the structure and metadata of programs rather than their behavior at runtime.
While we have reviewed many different feature extractors (see \autoref{sec:related}), we will focus on the Drebin, MaMaDroid, and MalScan feature set~\cite{ArpSHGR14,onwuzurike2019mamadroid,MalScan2020} in this work.
The Drebin feature set relies on permissions, API calls, and network addresses that can be extracted from input APKs, turning them into binary feature vectors.
The MaMaDroid feature set extracts the sequence of API calls retrieved by browsing the code. 
It then constructs a Markov chain to represent the transitions between API calls, grouped at varying levels of abstraction, such as API families or packages, to capture the APK’s behavioral patterns. Features derived from the Markov chain, such as transition probabilities, are used to train machine learning classifiers to distinguish between benign and malicious APKs.
The MalScan feature set extracts function call graphs of APKs based on static analysis and performs social-network-based centrality analysis to represent the semantic features of the graphs. The obtained feature vectors are fed to machine learning models for classification.

\mypar{Poisoning Attacks.}
These strategies manipulate the behavior of the targeted model by deliberately introducing tampered training samples. 
Poisoning attacks can be formalized as \emph{Denial of Service} (also known as \textit{availability}) and \textit{integrity} attacks~\cite{cina2023wild, biggio2012poisoning}.
A \emph{Denial of Service} (DoS) attack aims to subvert the overall accuracy at the test time of the model trained with poisoned training data. 
Consequently, the attack becomes noticeable only after the poisoned model has been deployed, not before. 
On the other hand, \emph{integrity} attacks seek to deform the victim model's performance on specific test samples while preserving the overall test-time accuracy. 
Attackers can employ various training data poisoning strategies, including \emph{clean-label}, \emph{dirty-label}, and \emph{label-flip} attacks. 
Specifically, \emph{clean-label} attacks indicate that the attacker can only tamper with the features of training samples without changing their labels. 
In contrast, in \emph{dirty-label} attacks, the attacker can manipulate both features and labels to poison training samples.
\emph{Label-flip} attacks only change the labels of training samples without changing their features.
\section{AndroVenom: Label Spoofing Attacks against Android Malware Detection}
\label{sec:androvenom}
\begin{figure}
    \centering
    \includegraphics[width=\linewidth]{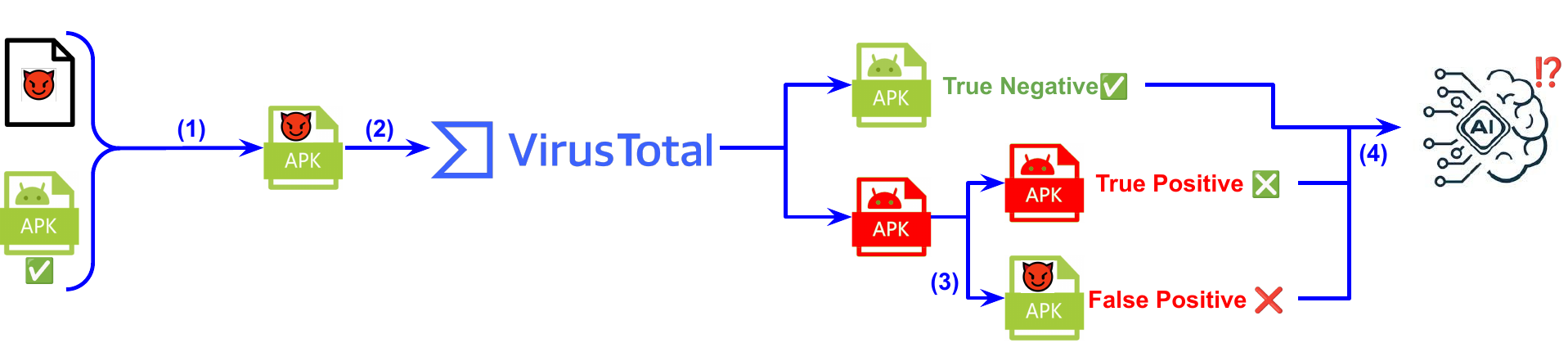}
    \caption{
    Workflow of AndroVenom:  (1) the attacker injects a malicious file into benign APKs and repacks them;
    (2) the attacker submits the modified benign APKs to VirusTotal, where they are now mislabelled as malware;
    (3) these mislabelled APKs are included in training sets of ML Android malware detectors;
    (4) the detectors are compromised after training.
    }\label{fig:workflow}
    % https://docs.google.com/drawings/d/15ZrMKf5t9PHAvHsSSfUX6TrsxLDWD4hPq-vCa85_Jkc/edit?usp=sharing
\end{figure}

We now describe AndroVenom (\rev{\autoref{fig:workflow}}), our methodology for poisoning training datasets used in ML-based malware classification via label spoofing. We first detail the selection process for identifying easy-to-spot malware samples intended for injection into benign APKs (\autoref{sec:sign_injection}). Subsequently, we outline the staging of the poisoning attacks (\autoref{sec:androvenom_poisoning}).

\rev{
Crucially, this attack does not target the internal learning mechanisms of AntiVirus (AV) engines themselves. Instead, it targets the labeling pipeline that feeds ML-based malware classifiers, which are often trained on crowd-sourced AV annotations.} \rev{
Our \textbf{threat model} does not assume that AV vendors directly retrain their primary detection engines on public datasets or VirusTotal submissions. Rather, the attack targets the downstream ML pipeline, where AV consensus (typically aggregated via platforms like VirusTotal) is treated as the ground truth for dataset curation. By coercing AV engines into producing incorrect labels through malicious signature injection, an attacker effectively contaminates the training data used for future ML-based detectors.}

\subsection{Label Spoofing: Coercing Mispredictions from AVs}
\label{sec:sign_injection}

\rev{We define \emph{label spoofing attacks} as attacks that aim to force an AV to misclassify a legitimate input APK as malicious and to attribute it to a specific malware family.}
This strategy is in stark contrast with the usual behavior of attackers that try to bypass AVs. 
\rev{To the best of our knowledge, we are the first to identify and analyze this counterintuitive behavior.}
In particular, we envision attackers that can easily pollute crowd-sourced datasets because of the blind trust attributed to annotations collected from AVs.
To do so, we first need to study how an attacker can manipulate benign APKs to achieve their goal.
Among all possible manipulations, in this work, we seek to inject malicious samples that satisfy different constraints to be considered effective: \emph{(i)} their inclusion must trigger as many AV detections as possible and be categorized as a specific malware family; \emph{(ii)} it should take significant engineering effort to filter out the presence of these malicious samples; and
\emph{(iii)} the injection location must be chosen accordingly to avoid impacting the output of feature extraction algorithms (we report our findings related to those questions in \autoref{tab:types}). 

\mypar{Detectable Malware Samples.} To satisfy the first constraint, we consider samples that trigger more than 20 AV detections on VirusTotal (see the \textit{Detection} column in \autoref{tab:types}). This conservative threshold was chosen to eliminate false positives, as prior research suggests that false detection rates drop significantly after 5 detections~\cite{zhu2020measuring}. Because we require these files to be consistently classified within a specific malware family, we must determine the consensus among potentially conflicting AV responses. We leverage the AVClass tool~\cite{sebastian2016avclass, sebastian2020avclass2} to resolve these labels, filtering out any samples for which a definitive malware family cannot be identified.

\mypar{Disguised as Common Files.} To minimize the risk of being flagged by manual analysis (the second constraint), we focus on malicious samples that share file types commonly found in benign APKs. For example, Unity-based games often contain Windows executables due to cross-platform deployment; consequently, a Windows library inside a mobile game archive may not appear suspicious. 
Our analysis utilizes the benign APK dataset proposed by Ruggia et al.~\cite{ruggia2022}, which comprises \goodwareCardinality benign APKs across 50 Google Play Store categories. 
Due to legal constraints, we used the hashes provided by the authors to download the APKs from VirusTotal. We then identified the \emph{ten most common MIME (Multipurpose Internet Mail Extensions) types} within these APKs and their likelihood of appearance (column \textit{Inclusion} in \autoref{tab:types}). We refined our selection of malicious samples by retaining only those that match these common MIME types. While thousands of samples met our criteria, we selected the smallest file for each MIME type to keep the payload footprint minimal.
\begin{table}
    \definecolor{lightgray}{gray}{0.92}
    \centering
    \resizebox{\linewidth}{!}{%
    \begin{tabular}{cccccc}
        \specialrule{1.2pt}{0pt}{0pt}
        \textbf{MIME Type}       & \textbf{Inclusion} & \textbf{Malicious File MD5}                     & \textbf{Detections} & \textbf{Family} & \textbf{Size [B]} \\ \hline
        \rowcolor{lightgray}
        text/xml                 & 100\%          & c4358d29533117779296b7e555aca54f & 29/62               & groooboor       & 316                   \\ %\hline
        app/octet-stream & 99.4\%          & 1ce7054c91ea5aca1e9ca75e97c7c1d8 & 33/62               & gnaeus          & 650                   \\ 
        \rowcolor{lightgray}
        image/gif                & 98.7\%          & 8575a8e68f927705a04e668fdc2246c1 & 32/62               & chopper         & 55                    \\ %\hline
        text/html                & 97.9\%          & b66b412c448f20d420ec83bb326b9f34 & 46/62               & scrinject       & 92                    \\ 
        \rowcolor{lightgray}
        text/plain               & 97.2\%          & a509796ddb1b0aa5e07f95b52dba7696 & 23/62               & smallasp        & 22                    \\ %\hline
        app/java-archive & 92.8\%          & b3ee9807df3e040d764a3bc795e01d5b & 22/64               & webshell        & 985                   \\ 
        \rowcolor{lightgray}
        app/json         & 91.6\%         &  1c3f8cd54127e869d4879a9b7499da3a & 21/62               & coinminer       & 1100                  \\ %\hline
        app/x-sharedlib  & 89.8\%         &  2506f9f2800c20da206776b0d43a5946 & 33/65               & lotoor          & 4505                  \\ 
        \rowcolor{lightgray}
        app/javascript   & 85.7\%         &  bca61a7bf8679fb23824f1c6c2d1a43c & 21/62               & scrinject       & 590                   \\ %\hline
        app/gzip         & 85.1\%         &  f0bea558da34ef8bd9561cb16caded27 & 31/62               & dloadr          & 782                   \\ 
        \specialrule{1.2pt}{0pt}{0pt}
        \end{tabular}}
    \caption{Top ten most common file types (MIME) and their likelihood of inclusion inside benign APKs, associated with a malicious sample of the same type.}
    \label{tab:types}
\end{table}

\begin{comment}
    text/xml
    https://www.virustotal.com/gui/file/a077501f8e57b4b1625869e05b09d3be672f5668445a5b74e27a0c1a7ed87884 29/62 groooboor

    app/octet-stream
    https://www.virustotal.com/gui/file/583e9d4df22034608cbbb7965d0af0c50c3e68a0b53ce14ca85e1420260525e3 33/62 gnaeus

    image/gif
    https://www.virustotal.com/gui/file/6b50e0c9432c88c3392862edcda8c634eb2832f796e4c0ead5afb0047fd690b8 32/62 chopper
    
    text/html
    https://www.virustotal.com/gui/file/7099fd770e5d795fdc4ae674ae86dfee6450632ddfaf441d210a3a0e10aca80b 46/62 scrinject
    
    text/plain
    https://www.virustotal.com/gui/file/95a46cba579c637b20b5e5c94a2029716dc6cc75b8e346c4cdccfb47c0c02c3d/ 23/62 smallasp
    
    app/java-archive
    https://www.virustotal.com/gui/file/ec31a1b9d3b5672137349a35719e2f595394a90f3978d60ef5ffe7900763ac00/ 22/64 webshell
    
    app/json
    https://www.virustotal.com/gui/file/4c9f6dc32415815c4be506f93f9cac7ac2996009461dadce79dfb08f1a45c59e 21/62 coinminer
    
    app/x-sharedlib
    https://www.virustotal.com/gui/file/2d2893abf8c1ff48cf070bd182353913612127c494d03442365a7c980de35f3f/ 33/65 lotoor
    
    app/javascript
    https://www.virustotal.com/gui/file/d06016d1e4f2cf27725265b6de5527bb824ea9f4c38d5fd89b2e019a27e3763f 21/62 scrinject

    app/gzip
    https://www.virustotal.com/gui/file/93f09128e5801c722faaa63647c883fd48739fbfbeae7f543151d1754294ce41 31/62 dloadr
    
\end{comment}

\mypar{Invisible Injection.} To satisfy the third constraint, we identified locations within the APK that are typically overlooked by feature extractors. We analyzed state-of-the-art extractors summarized in \autoref{Table:related poisoning attacks}, focusing on {Drebin}, {MaMaDroid}, and {MalScan} (see \autoref{sec:background}). Our analysis reveals that these malware feature extractors consistently ignore the content of the \emph{res} folder. Conversely, we exclude \texttt{AndroidManifest.xml} and DEX files, as they contain the permissions and bytecode analyzed by nearly all established feature extraction algorithms. By satisfying these constraints with the 10 malicious samples highlighted in \autoref{tab:types}, we demonstrate that attackers can use Apktool to unpack a benign APK, inject the malicious files into the \emph{res} folder, and repack it. This confirms the feasibility of \textbf{RQ1}.

\subsection{Poisoning Attacks against ML Malware Detectors}
\label{sec:androvenom_poisoning}
We leverage label spoofing to execute poisoning attacks, which are formalized as an optimization problem:
\begin{equation}
\tilde{D} \in \max_{D} \mathcal{L}(D_{ts}, f^\star) \quad s.t. \quad f^\star \in \min_{f \in \mathbf{H}} L(D_{tr} \cup D)\label{eq:poisoning_untargeted}
\end{equation}
where $\mathbf{H}$ denotes the hypothesis space of classifiers $f$; $\mathcal{L}$ and $L$ are loss functions quantifying errors at test and training time, respectively; $D_{tr}$ and $D_{ts}$ are the training and test sets; $\tilde D$ is the set of poisoned data; and $f^\star$ is the model trained on the union of regular and poisoned data. When $D_{ts}$ represents the entire test set, \autoref{eq:poisoning_untargeted} describes an availability or \emph{Denial of Service} (DoS) attack. In our realistic threat model, the attacker uses label spoofing to flip labels from benign to malicious by repackaging the application content. This scenario differs from traditional clean-label or label-flip settings because the attacker does not assign the label directly; rather, they exploit the AV's internal logic to generate the desired label, leveraging the implicit trust practitioners place in these systems.

\mypar{Maiming the Reputation of Legitimate APKs.} The formulation in \autoref{eq:poisoning_targeted} can also represent targeted misclassifications:
\begin{equation}
\tilde{D} \in \max_{D} \mathcal{L}(\vct x_t, f^\star) \quad s.t. \quad f^\star \in \min_{f \in \mathbf{H}} L(D_{tr} \cup D)\label{eq:poisoning_targeted}
\end{equation}
In this case, the attacker maximizes the error for a specific target sample $\vct x_t$ (a benign APK). The resulting model maintains general performance but misclassifies the chosen point. To trigger this specific misclassification, we use clones of the targeted APK, a strategy inspired by~\cite{aghakhani2021bullseye}. We augment label spoofing by injecting API calls into the APK bytecode, guarded by opaque predicates (logical statements that are always true or false). This ensures that the original functionality remains intact while static feature extractors are tricked into processing the injected calls. By distributing these clones and forcing AVs to label them as malicious, we populate the feature space with samples that surround the target benign APK, forcing a misprediction at test time \emph{only} for that specific application. As shown in \autoref{sec:targeted}, in some configurations, only five clones are required to successfully flip a detection.

\section{Experiments}\label{sec:experiments}
We now conduct an extensive experimental analysis to demonstrate the feasibility of label spoofing (\autoref{subsec:feasability}) and its impact on ML models in both untargeted and targeted scenarios (\autoref{sec:untargeted} and \autoref{sec:targeted}). Finally, we show that existing countermeasures may fall short, not only in detecting these attacks but also in maintaining reliable predictive performance when deployed (\autoref{sec:countermeasure}).

\subsection{Experiment Setup}
\label{subsec:exp_setup}

\mypar{Dataset.} We replicate the dataset proposed by Ruggia et al.~\cite{ruggia24unmasking}, which consists of malicious and benign APKs collected between September 2022 and April 2023. Due to legal restrictions, the authors provided only the APK hashes; we subsequently downloaded the corresponding files from VirusTotal to ensure our data source remained consistent with the original study. Our collection includes a \emph{Malware Dataset} of \malwareCardinality samples across 196 families (ranging from 100 to 114 samples per family) and a \emph{Goodware Dataset} of \goodwareCardinality benign APKs spanning 50 Google Play Store categories.We conduct our experiments across two settings: \emph{(i)} a \emph{large-scale} experiment using a dataset \largedata of 40K samples, equally partitioned between the \emph{Malware} and \emph{Goodware} datasets; and \emph{(ii)} a \emph{small-scale} experiment using a dataset \smalldata of 1K samples. The latter comprises 500 benign and 500 malicious APKs, sampled from \largedata to include 10 benign samples from each Google Play Store category and 2 to 3 malicious samples from each family. In both settings, we maintain a balanced class ratio to minimize the impact of class imbalance on classification performance, thereby highlighting the specific accuracy degradation caused by label-spoofed training samples.

\begin{figure}[t]
  \centering
   \includegraphics[width=\linewidth]{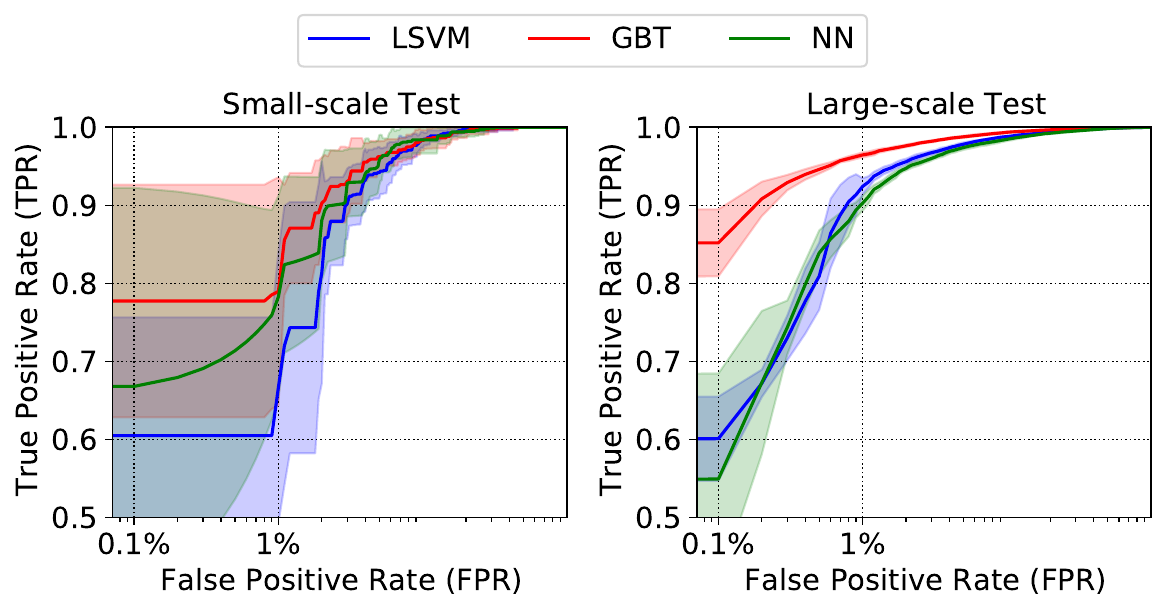} %width=.85
    \caption{ROC curves of three classifiers (LSVM, GBT, NN) using Drebin features on the small-scale and large-scale test.}
    \label{fig:ROC_0p_3models}
\end{figure}

\begin{figure}[t]
  \centering
   \includegraphics[width=\linewidth]{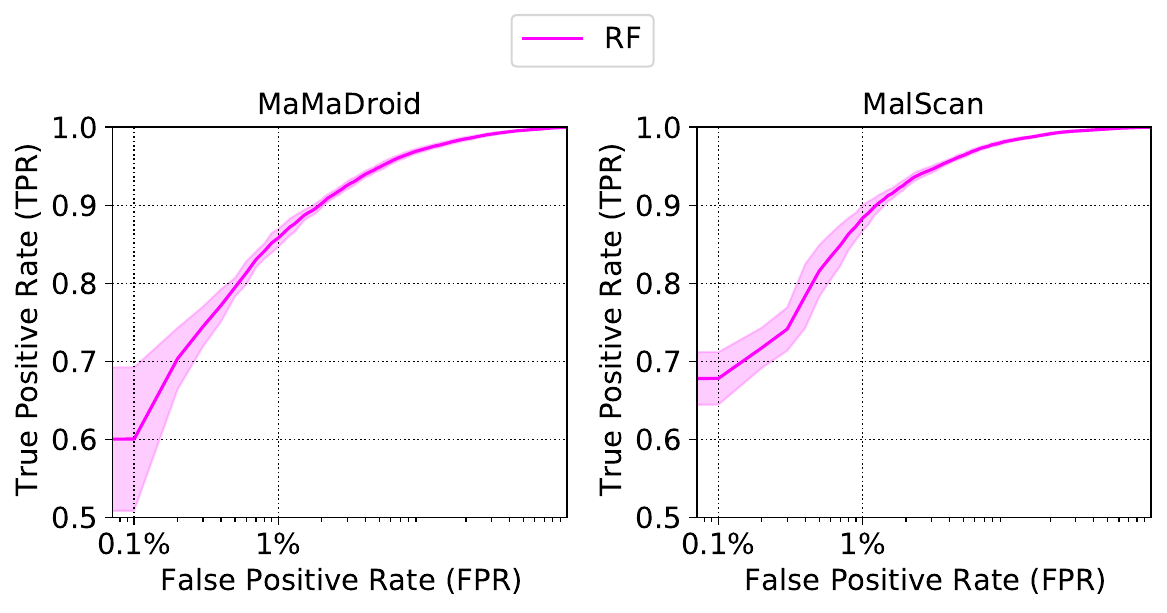} %width=.85
    \caption{ROC curves of RF using MaMaDroid and MalScan features on the large-scale test.}
    \label{fig:ROC_40K_0p_RF_mamadroid_malscan}
\end{figure}

\mypar{Feature Extractors and Model Architectures.} We evaluate three prominent feature extractors, Drebin, MaMaDroid, and MalScan~\cite{ArpSHGR14, onwuzurike2019mamadroid, MalScan2020}, paired with various ML-based malware classification models. 
Following established literature~\cite{anderson2018ember, ArpSHGR14, dambra2023decoding}, we pair the Drebin feature extractor with a Linear Support Vector Machine (LSVM) and Gradient Boosted Trees (GBT). 
We set the LSVM regularization parameter to $\lambda=1$, while the GBT is configured with 300 trees for \smalldata and 3,000 trees for \largedata. 
Additionally, we implement a Neural Network (NN) featuring a 5-layer feed-forward architecture (with 24000, 240, 120, 60, and 1 neurons, respectively) and Sigmoid activation functions to generate a final probability of malice. 
To manage the high dimensionality of Drebin feature vectors, which reach 732,800 features in our dataset, we utilize Random Projections~\cite{Li2006kdd} to reduce the input dimensionality to 24,000 before feeding it into the NN. The NN is trained for 5,000 epochs using the Adam optimizer~\cite{kingma2017adam} to minimize cross-entropy loss. 
Similarly, we evaluated MaMaDroid and MalScan with several models suggested in prior work~\cite{onwuzurike2019mamadroid, MalScan2020}. 
However, based on empirical performance, we ultimately selected a Random Forest (RF) architecture with 300 trees for these extractors. We apply Drebin features exclusively in the small-scale experiment and utilize all three feature types in the large-scale experiment. For both \smalldata and \largedata, the data is split into 80\% training and 20\% test sets. 
To account for training uncertainty, we repeat the train-test split process 10 times. We report the mean and standard deviation of the resulting metrics, visualizing the Receiver Operating Characteristic (ROC) curves for Drebin-based models in \autoref{fig:ROC_0p_3models}, and for MaMaDroid and MalScan in \autoref{fig:ROC_40K_0p_RF_mamadroid_malscan}.

\mypar{Performance Metrics.} 
To evaluate the performance of our classifiers, we report two primary metrics: \emph{(i)} the False Positive Rate (FPR), which quantifies the volume of false alarms generated; and \emph{(ii)} the True Positive Rate (TPR), which measures the proportion of correctly identified malware samples. For each classifier, we calculate the classification thresholds required to maintain fixed FPR levels of 0.1\% and 1\%. Finally, we utilize the Attack Success Rate (ASR) to quantify the efficacy of integrity attacks, defined as the frequency with which the attack successfully triggers a misclassification of the targeted benign APK.

\mypar{Countermeasure.} We also consider a scenario in which the ML service provider adopts defense mechanisms to mitigate potential poisoning noise within the training data. Specifically, we evaluate the Deep k-NN defense proposed by Peri et al.~\cite{Peri2020ECCV}, which is designed to identify mislabeled training samples. This method performs anomaly detection by examining the labels of the $k$-nearest neighbors~\cite{dasarathy1991nearest} for each sample in the feature space. We demonstrate the effectiveness of this countermeasure against both Denial of Service and targeted misclassification attacks in \autoref{sec:countermeasure}.

\subsection{Feasibility of Label Spoofing Attacks}
\label{subsec:feasability}

We demonstrate the feasibility of label spoofing attacks by utilizing the malicious samples selected in \autoref{sec:androvenom} and investigating: \emph{(i)} whether their injection via repackaging is viable; \emph{(ii)} the potential side effects of such perturbations; and \emph{(iii)} whether AV engines are effectively coerced into erroneous misclassifications.

\mypar{Injection through Repackaging.}
Inserting a file into an APK requires a repackaging process, as described in \autoref{sec:background}. This process may fail due to limitations in the Apktool library or defensive mechanisms implemented within the APK to prevent tampering~\cite{rastogi2016android}. Therefore, we first assessed the success rate of repackaging by applying Apktool to all samples in our \emph{Goodware Dataset}. Our results show that 97.7\% (\repackagedCardinality/\goodwareCardinality) of the benign APKs were successfully repackaged without errors. We refer to this subset as the \emph{Repack-able Goodware Dataset}.

To comply with Android OS requirements and ensure the APKs remain installable, we digitally signed each sample using a custom-generated certificate. To maintain ethical standards and prevent misuse, we included metadata within the certificate explicitly stating it was generated for scientific research purposes.

\mypar{Side Effects of the Injection.}
We next quantified the side effects resulting from the injection process. To do so, we constructed a \emph{Spoofed Dataset} by repackaging each APK in the \emph{Repack-able Goodware Dataset} with a randomly selected malicious sample from those listed in \autoref{tab:types}. As specified in \autoref{sec:androvenom}, the malicious file is inserted into the \texttt{res} folder before the APK is repackaged.

We measured the size variation of these repackaged APKs relative to their original versions to determine if the process results in noticeably larger archives. Our analysis indicates that the impact on file size is negligible; on average, the APK size increased by only 2--3\%, with a median increase of approximately 1.4\%.

\subsection{Influence on the Decisions of AVs.}

\rev{In this section, we designed two experiments to balance realism with ethical responsibility. The first experiment utilizes VirusTotal to demonstrate the feasibility of label spoofing on a real-world AV aggregation platform. However, since the repeated submission of synthetic malware can pollute publicly scraped datasets, we limited the number of VirusTotal uploads. To evaluate the attack mechanism at scale without further contamination, we designed a second experiment that replicates VirusTotal’s behavior within a controlled environment using multiple commercial Android AV products. This dual-pronged approach allows for a large-scale study of label spoofing while avoiding additional public data pollution.}

\rev{\mypar{First Experiment.} We randomly sampled \synthVTCardinality APKs from the \emph{Spoofed Dataset} and uploaded them to VirusTotal.} While this sample size is relatively small compared to our primary datasets, and despite the fact that uploading synthetic malware for research is generally tolerated~\cite{aonzo2020obfuscapk, zhu2020measuring, huang2024exploring, botacin2023gpthreats}, we deliberately restricted the volume of submissions to minimize the risk of contaminating public datasets. For transparency, we added a comment to each VirusTotal submission page stating that the APK was synthetic malware generated for research purposes. \rev{All \synthVTCardinality\ APKs were flagged as malicious, with an average detection count of $13 \pm 2$ out of the 62 AV engines available at the time of submission (a detection rate of $21 \pm 2.9\%$). This detection range falls within the stable thresholds identified by Zhu et al.~\cite{zhu2020measuring}, which are largely insensitive to short-lived label fluctuations and engine-level noise, thereby indicating a high-confidence malicious classification.} Furthermore, applying AVClass~\cite{sebastian2020avclass2} to the reports consistently attributed the samples to the same malware family as the injected malicious component.

To confirm that these detections were triggered by the injected payloads rather than the repackaging process itself, we conducted a second control experiment. We submitted the entire set of 20,284 APKs from the \emph{Repack-able Goodware Dataset} to VirusTotal. \rev{In this case, \emph{none} of the samples was flagged. This serves as a negative control, validating that repackaging alone does not induce false positives and that the previous detections are exclusively attributable to the injected malicious artifacts.} Finally, as an ethical precaution, we requested the deletion of our \synthVTCardinality synthetic samples from VirusTotal; our request was promptly accepted.

\rev{\mypar{Second Experiment.} To avoid further VirusTotal pollution, we sought to replicate its behavior in a local environment. However, fully replicating VirusTotal is infeasible, as it relies on proprietary, cloud-based AV engines that cannot be deployed locally. Consequently, we} selected the 14 Android AVs evaluated in July 2025 by AV-Test~\cite{avtestJul25}, representing the most established solutions on the market. We prepared 14 Android emulators on a physical server equipped with four Intel Xeon Platinum 8160 CPUs (2.10 GHz, 96 cores total) and 512 GB of RAM. Each emulator was configured with one of the 14 AVs and its corresponding signature database. Subsequently, the server was disconnected from the Internet to prevent the emulators from communicating with vendor servers, ensuring no samples were inadvertently uploaded.

We observed that real-time protection functioned in only 9 out of the 14 AVs without Internet access. This limitation is well-documented and motivates the use of on-device ML models~\cite{aonzo2017low,peng2022lightweight}. Using a known Joker malware sample, we confirmed that upon installation, the AV issues a warning notification that can be programmatically retrieved via \texttt{dumpsys}~\cite{dumpsys} through the Android Debug Bridge (ADB).

In this setup, we successfully installed 19,067 out of \repackagedCardinality\ (93\%) APKs from the \emph{Spoofed Dataset} across the 9 functioning emulators. The primary cause of installation failure was architecture incompatibility (ARM-only APKs on x86 emulators). Notably, \emph{all} successfully installed APKs were flagged as malicious by at least one AV. However, the average detection rate was lower than in the VirusTotal experiment ($2 \pm 0.02\%$). This reduction is expected, as mobile AV versions typically lack the full signature sets available to cloud-based aggregators, likely due to local storage constraints, and often rely on remote lookups \rev{that were disabled in our experiment to prevent polluting the companies' datasets.}

Despite these environmental differences, the results remain representative of realistic deployment conditions and support \textbf{RQ2} by demonstrating that label spoofing attacks are both feasible and effective in practical scenarios.

\rev{
\subsection{Case Studies}
\mypar{Case Study 1.}
We consider a benign APK belonging to the Games category, originally undetected by all AV engines on VirusTotal. 
We injected the 55-byte \texttt{image/gif} malware sample of the known Chopper (see \autoref{tab:types}) into the \texttt{res/drawable} directory. 
After repackaging and resigning, the APK was flagged as malicious by 22/62 AVs on VirusTotal. 
AVClass consistently attributed the detection to the Chopper family. 
Re-submitting the same APK without the injected file resulted in zero detections, confirming that the alert was exclusively caused by the injected artifact.
}

\rev{
\mypar{Case Study 2.}
We injected the malicious 92-byte \texttt{text/html} file into \texttt{assets/www/index.html}, a path commonly used by Apache Cordova hybrid
apps, corresponding to the Scrinject malware family, into an otherwise benign productivity APK. 
On VirusTotal, 22/62 AVs flagged the modified APK as malicious, while it did not flag the same APK without the injected file.
Interestingly, when evaluating the APK in an on-device/emulator setting, 3 out of 9 AV products detected the sample, although none of their corresponding engines on VirusTotal had flagged it in the previous experiment.
These results highlight that identical injected artifacts can trigger heterogeneous detection behaviors across AV engines and deployment settings, yet still consistently succeed in inducing malicious labels.
}

\subsection{Results of DoS Attacks}
\label{sec:untargeted}
We now show how AndroVenom can induce Denial of Service (DoS) against ML models trained spoofed data on both \smalldata and \largedata.
For each train-test split, we randomly select a percentage $p=1\%, 2\%, 5\%, 10\%, 20\%$ of the benign APKs in the training set to produce misclassifications, amounting to 8, 16, 40, 80, 160 APKs in \smalldata, and 328, 657, 1643, 3287, 6575 APKs in \largedata. 
Since we have already proved the feasibility of label spoofing through repackaging, and since we do not want to pollute the data sources of VirusTotal, we will \emph{simulate} our attacks, without sending any more samples to be analyzed. 
%We show the drop in terms of TPR and FPR using Drebin features for the small-scale test and the large-scale test in \autoref{fig:ROC_1K40K_untargeted_3models}, and using MaMaDroid features for the large-scale test in \autoref{fig:ROC_40K_untargeted_RF_legend}. 

\begin{figure}[t]
  \centering
  \includegraphics[width=\linewidth]{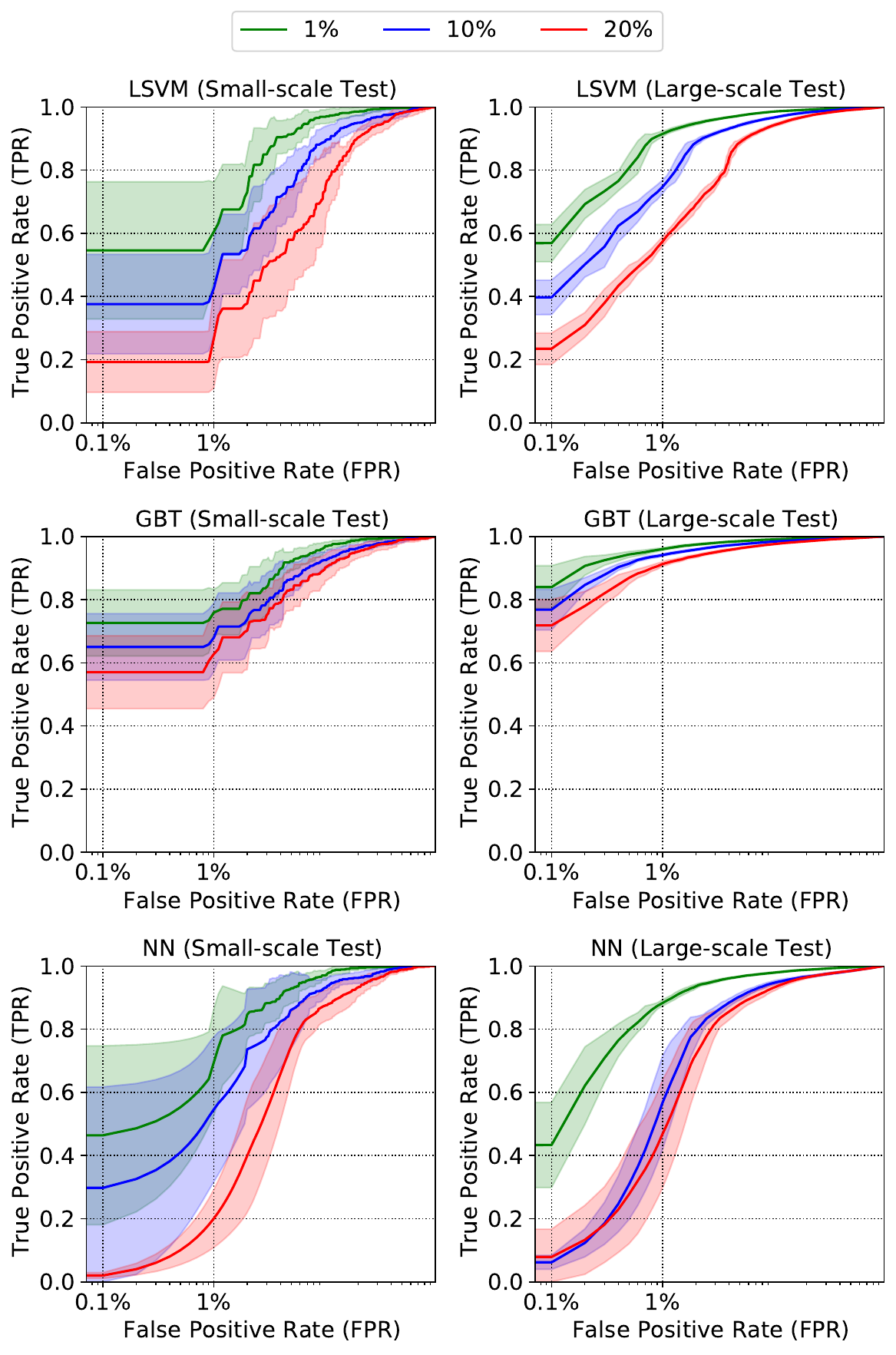}
  \caption{ROC curves of LSVM, GBT, NN using Drebin features on the small-scale and large-scale test with the poisoning ratio 1\%, 10\%, and 20\%.}
  \label{fig:ROC_1K40K_untargeted_3models}
\end{figure}

\begin{figure}[t]
  \centering
  \includegraphics[width=\linewidth]{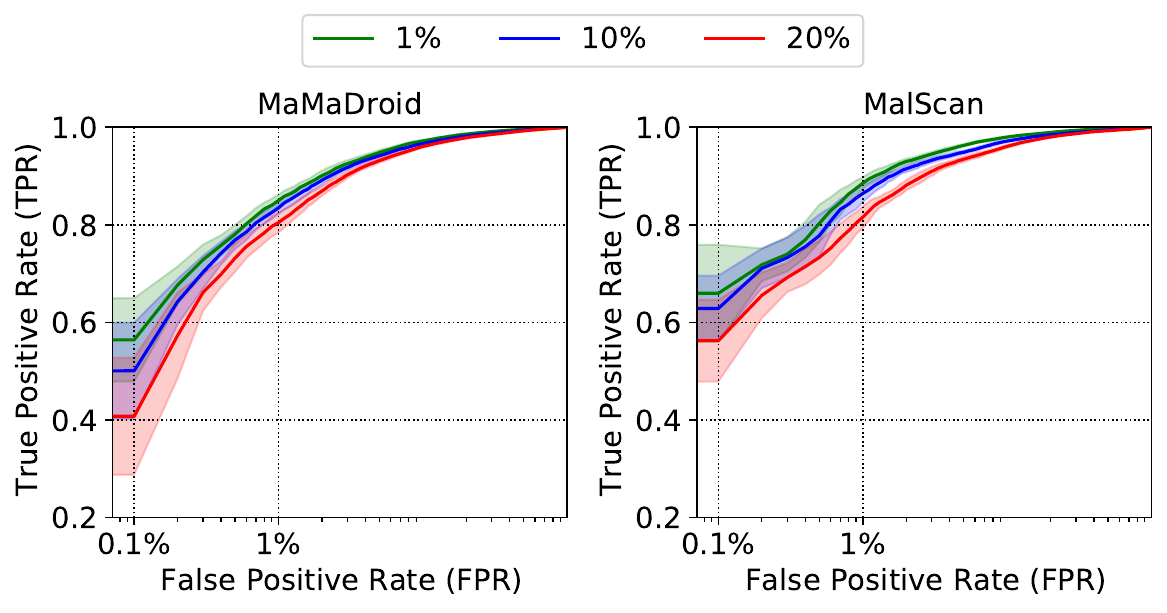}
  \caption{ROC curves of RF using MaMaDroid and MalScan features on the large-scale test with the poisoning ratio 1\%, 10\%, and 20\%.}
  \label{fig:ROC_40K_untarget_mamadroid_malscan}
\end{figure}

\noindent\textbf{Effect on Small-scale Dataset.} 
We show the efficacy of DoS against Drebin-based models in \autoref{fig:ROC_1K40K_untargeted_3models}, highlighting a consistent drop in TPR with the increasingly larger poisoning ratio for all three classifiers.
In particular, as shown in \autoref{Table:TPR_1K_poison_3models}, with only 1\% of label-spoofed APKs (8 samples out of 800), the TPR of the poisoned classifier drop by 5.9\%, 5.1\%, 20.4\% for LSVM, GBT, and NN respectively, compared to the poison-free baseline when FPR is 0.1\%.
We perform Mann-Whitney U tests to compare TPR values at consecutive poisoning ratio levels—such as 0\% vs. 1\% and 1\% vs. 2\%, for FPR thresholds of 0.1\% and 1\%. In most cases, the test rejects the null hypothesis (p $<$ 0.05), indicating a statistically significant drop in TPR as the poisoning ratio is increased.
Among the three classifiers, NN is the most vulnerable to the DoS attack, with a drop of TPR below 50\% at 0.1\% FPR for all attacks that we evaluate in the experiments.

Although LSVM preserves relatively more classification accuracy when the poisoning ratio is low, its TPR still has a large decrease when the poisoning ratio rises to 5\%, (40 samples out of 800). 
Among all the three ML models, GBT is the most robust classifier against DoS attacks, with only a 16.3\% TPR drop at 1\% FPR with 20\% of label-spoofed samples. As GBT uses an ensemble of decision trees, each tree minimizes the errors of the previous ones. As a result, GBT reduces the impact of DoS attacks, benefiting from multiple boosted training iterations.

\begin{table*}[t]
  \centering
  \footnotesize
  \caption{Mean TPR with standard deviation of models trained with Drebin features on \smalldata against DoS attacks at different poisoning ratios. We report performances $0.1\%$ and $1\%$ FPR. All numbers are in percentage.}
  \label{Table:TPR_1K_poison_3models}
\definecolor{lightgray}{gray}{0.92}
  \resizebox{0.9\textwidth}{!}{ 
  % \small
  \begin{tabular}{cccccccc}    
   %\cline{3-8} 
   \specialrule{1.2pt}{0pt}{0pt}
   & & \multicolumn{6}{c}{\textbf{Poisoning Ratio}} \\
   %\cline{3-8} 
   &   &   \textbf{0}      &   \textbf{1}      & \textbf{2}   & \textbf{5}  & \textbf{10}  & \textbf{20}       \\
\hline
\rowcolor{lightgray}
   \multicolumn{1}{c}{\multirow{2}{*}{\cellcolor{white}\textbf{LSVM}}} & \textbf{FPR=0.1} &    $60.5 \pm 15.1$    &  $54.6 \pm 21.7$     &   $54.3 \pm 22.0$   &   $47.5 \pm 17.3$  &  $37.6 \pm 15.7$   &  $19.2 \pm 9.6$      \\ 
   
   \multicolumn{1}{c}{}  & \textbf{FPR=1}   &     $66.6 \pm 17.4$     &    $60.5 \pm 15.7$     &  $61.2 \pm 18.1$  &    $50.5 \pm 13.4$  &    $42.5 \pm 15.3$  &   $26.6 \pm 15.6$     \\ 
   \hline
   \rowcolor{lightgray}
   \multicolumn{1}{c}{\multirow{2}{*}{\cellcolor{white}\textbf{GBT}}} & \textbf{FPR=0.1} &    $77.7 \pm 14.9$    &  $72.6 \pm 10.5$     &   $69.7 \pm 10.5$   &   $71.7 \pm 7.9$  &  $65.0 \pm 10.6$   &  $57.0 \pm 11.5$      \\ 
 
   \multicolumn{1}{c}{}  & \textbf{FPR=1}   &     $79.0 \pm 14.6$     &    $75.9 \pm 7.4$     &  $74.9 \pm 7.7$  &    $76.5 \pm 8.0$  &    $67.9 \pm 10.9$  &   $62.7 \pm 13.8$     \\ 
   \hline
   \rowcolor{lightgray}
   \multicolumn{1}{c}{\multirow{2}{*}{\cellcolor{white}\textbf{NN}}} & \textbf{FPR=0.1} &    $66.8 \pm 25.4$    &  $46.4 \pm 28.3$     &   $32.3 \pm 27.8$   &   $33.5 \pm 35.8$  &  $29.8 \pm 32.0$   &  $2.0 \pm 0.9$      \\ 
 
   \multicolumn{1}{c}{}  & \textbf{FPR=1}   &     $78.3 \pm 12.9$     &    $69.5 \pm 16.9$     &  $69.2 \pm 10.7$  &    $63.6 \pm 23.4$  &    $54.6 \pm 23.0$  &   $20.0 \pm 9.2$     \\ 
\specialrule{1.2pt}{0pt}{0pt}
  \end{tabular}
  }
\end{table*}

\noindent\textbf{Effect on Realistic Large-scale Dataset.} 
The best malware classification performance of ML models, trained on the three feature representations and on the large-scale dataset, are increased, thanks to the abundance of training data. However, we observe the same trend highlighted from the small-scale experiment: \emph{TPR of malware classification presents sharp decreases in the presence of attacks}.

We show ROC curves and results at different poisoning ratios against Drebin-based, MaMaDroid-based, and MalScan-based models in \autoref{fig:ROC_1K40K_untargeted_3models}, \autoref{fig:ROC_40K_untarget_mamadroid_malscan}, \autoref{Table:TPR_40K_poison_3models}.
In the case of Drebin, when the poisoning ratio is over 2\% for NN and 5\% for LSVM, TPR drops significantly compared to the poison-free baseline. 
With 0.1\% FPR, the label spoofing attack on 1643, 3287, and 328 benign APKs out of 32K training samples are already sufficient to cause a 10\% drop of TPR for LSVM, GBT, and NN respectively.
For MaMaDroid features, we observe a 7\% drop of TPR at 0.1\% FPR, but unable to reduce further the performance similar to GBT.
For MalScan features, TPR drops less than 2\% at both 0.1\% and 1\% FPR while the poisoning ratio is less than 5\%, demonstrating its robustness compared to other features. Nevertheless, TPR decreases evidently after 5\% poisoning ratio.
\begin{table}[t]
  \scriptsize
  \centering
  \definecolor{lightgray}{gray}{0.92}
  \caption{Mean FPR of models trained with Drebin features on \largedata against DoS attacks at different poisoning ratios. We report performances at $95\%$ TPR. All numbers are in percentage.}
  \label{Table:FPR_40K_poison_3models}
  \resizebox{0.9\linewidth}{!}{ 
  \begin{tabular}{ccccccc}    
   %\cline{2-7} 
   \specialrule{1.2pt}{0pt}{0pt}
   & \multicolumn{6}{c}{\textbf{Poisoning Ratio}} \\
   %\cline{2-7} 
   &   \textbf{0}      &   \textbf{1}      & \textbf{2}   & \textbf{5}  & \textbf{10}  & \textbf{20}       \\
   \hline
   \multicolumn{1}{c}{\textbf{LSVM}} &    $1.7$    &  $2.2$     &   $2.4$   &   $3.6$  &  $6.0$   &  $12.0$      \\ 
   \rowcolor{lightgray}
   \multicolumn{1}{c}{\textbf{GBT}} &    $0.6$    &  $0.7$     &   $0.8$   &   $1.0$  &  $1.4$   &  $3.0$      \\ 
   
   \multicolumn{1}{c}{\textbf{NN}}  &    $2.1$    &  $2.8$     &   $3.7$   &   $6.9$  &  $11.1$   &  $13.9$      \\ 
   %\hline
   \specialrule{1.2pt}{0pt}{0pt}
  \end{tabular}
  }
\end{table}

\begin{table}[t]
  \scriptsize
  \centering
  \definecolor{lightgray}{gray}{0.92}
  \caption{Mean FPR of RF trained with MaMaDroid and MalScan features on \largedata against DoS attacks at different poisoning ratios. We report performances at $90\%$ TPR. All numbers are in percentage.}
  \label{Table:FPR_40K_poison_2models_mamadroid_malscan}
  \resizebox{0.9\linewidth}{!}{ 
  \begin{tabular}{ccccccc}    
   %\cline{2-7}  
   \specialrule{1.2pt}{0pt}{0pt}
   & \multicolumn{6}{c}{\textbf{Poisoning Ratio}} \\
   %\cline{2-7}  
   &  \textbf{0}      &   \textbf{1}      & \textbf{2}   & \textbf{5}  & \textbf{10}  & \textbf{20}       \\
   \hline
     \rowcolor{lightgray}
     \multicolumn{1}{c}{\textbf{MaMaDroid}}  &   $ 1.9 $    &  $ 2.0 $     &   $ 2.0 $   &   $ 2.1 $  &  $ 2.3 $   &  $ 2.9 $      \\ 

    \multicolumn{1}{c}{\textbf{MalScan}}  &    $ 1.2 $    &  $ 1.2 $     &   $ 1.2 $   &   $ 1.3 $  &  $ 1.6 $   &  $ 2.6 $      \\ 
    \specialrule{1.2pt}{0pt}{0pt}
  \end{tabular}
  }
\end{table}

% \begin{table}[t]
%   \scriptsize
%   \centering
%   \caption{Mean FPR of RF trained with MaMaDroid features on \largedata against DoS attacks at different poisoning ratios. We report performances at $90\%$ TPR. All numbers are in percentage.}
%   \label{Table:FPR_40K_poison_2models_mamadroid}
%   \resizebox{0.8\linewidth}{!}{ 
%   \begin{tabular}{|c|c|c|c|c|c|}    
%    \hline 
%     \multicolumn{6}{|c|}{\textbf{Poisoning Ratio}} \\
%    \hline 
%       \textbf{0}      &   \textbf{1}      & \textbf{2}   & \textbf{5}  & \textbf{10}  & \textbf{20}       \\
%    \hline
%        $ 1.9 $    &  $ 2.0 $     &   $ 2.0 $   &   $ 2.1 $  &  $ 2.3 $   &  $ 2.9 $      \\ 
%    \hline
%   %  \multicolumn{1}{|c|}{\textbf{NN}}  &    $ 5.5 $    &  $ 5.6 $     &   $ 5.7 $   &   $ 6.1 $  &  $ 6.9 $   &  $ 8.7 $      \\ 
%   %  \hline
%   \end{tabular}
%   }
% \end{table}
We also quantify the effect of poisoning by highlighting the increment of FPRs.
To do so, we freeze TPR to 95\% and observe the variation of FPR with an increasingly larger poisoning ratio. 
For Drebin-based classification, as we can see in \autoref{Table:FPR_40K_poison_3models}, FPR of GBT rises from 0.6\% to 1.0\%, 1.4\% and 3.0\% respectively with the poisoning ratio as 5\%, 10\% and 20\%. 
We can observe similar trends for LSVM and NN. 
The Pearson correlation coefficients between FPR and the poisoning ratios for LSVM, GBT, and NN are all larger than 0.95, with p-values less than 0.05. 
This suggests a strong proportionality between the increase of the poisoning ratio and the rise of FPR. 
Besides, we show the FPR table for RF using MaMaDroid and MalScan features in \autoref{Table:FPR_40K_poison_2models_mamadroid_malscan} by freezing the TPR value to 90\%.
For MaMaDroid, FPR rises from 1.9\% to 2.1\%, 2.3\% and 2.9\% corresponding to the poisoning ratio of 5\%, 10\% and 20\% respectively. 
For MalScan, FPR rises from 1.2\% to 1.3\%, 1.6\% and 2.6\% corresponding to the poisoning ratio of 5\%, 10\% and 20\% respectively.
The Pearson correlation coefficients between FPR and the poisoning ratios for RF using MaMaDroid and MalScan features are larger than 0.95, with p-values less than 0.005.
This alignment with the attack objectives of AndroVenom underscores the effectiveness of the applied DoS attack methodology.
 
\mypar{Final Remarks on DoS Attacks.}
Given the results we presented, we are able to answer \textbf{RQ3}. Regardless of the size over training set, AndroVenom influences consistently ML-based malware classifiers. 
Security entities may use a small or large collection of file samples to build their ML-based solutions for malware classification. 
Nevertheless, an attacker may exploit the AV engine-based annotation process and inject an excessively small number of misannotated training samples into the training set to cause drastic utility loss of the learned classifier. 
This vulnerability existing in the AV-based annotation technique raises a severe trustworthy concern for the current practices of ML-based malware classification.

%the untargeted label poisoning attack has a significant negative impact on the classification performance of the three classifiers in a small-scale test. 
%To answer \textbf{RQ2}, we find that 40, 80, and 8 poisoned benign samples can prevent LSVM, GBT, and NN from functioning properly in $\mathcal{D}_{\text{1K}}$.

% Table:TPR_40K_poison_3models
\begin{table*}[t]
  \centering
  \footnotesize
  \caption{Mean TPR with standard deviation of models trained with Drebin, MaMaDroid, and MalScan features on \largedata against DoS attacks at different poisoning ratios. We report performances  $0.1\%$ and $1\%$ FPR. All numbers are in percentage.}
  \label{Table:TPR_40K_poison_3models}
  \definecolor{lightgray}{gray}{0.92}
  \resizebox{0.9\linewidth}{!}{ 
  %\small
  \begin{tabular}{cccccccc}    
   %\cline{3-8} 
   \specialrule{1.2pt}{0pt}{0pt}
   & & \multicolumn{6}{c}{\textbf{Poisoning Ratio}} \\
   %\cline{3-8}
   &   &   \textbf{0}      &   \textbf{1}      & \textbf{2}   & \textbf{5}  & \textbf{10}  & \textbf{20}       \\
   \hline
\rowcolor{lightgray}  
   \multicolumn{1}{c}{\multirow{2}{*}{\cellcolor{white}\textbf{Drebin LSVM}}} & \textbf{FPR=0.1} &    $60.1 \pm 5.4$    &  $56.9 \pm 5.9$     &   $55.1 \pm 4.3$   &   $48.1 \pm 8.4$  &  $39.7 \pm 5.5$   &  $23.4 \pm 5.0$      \\ 
   
   \multicolumn{1}{c}{}  & \textbf{FPR=1}   &     $92.4 \pm 1.1$     &    $91.4 \pm 1.0$     &  $90.7 \pm 0.9$  &    $86.5 \pm 3.3$  &    $74.6 \pm 2.1$  &   $57.3 \pm 1.6$     \\ 
   \hline
 \rowcolor{lightgray}
   \multicolumn{1}{c}{\multirow{2}{*}{\cellcolor{white}\textbf{Drebin GBT}}} & \textbf{FPR=0.1} &    $85.2 \pm 4.3$    &  $83.9 \pm 6.8$     &   $81.9 \pm 6.1$   &   $78.1 \pm 6.5$  &  $76.8 \pm 6.4$   &  $71.8 \pm 8.2$      \\ 

   \multicolumn{1}{c}{}  & \textbf{FPR=1}   &     $96.5 \pm 0.3$     &    $96.0 \pm 0.4$     &  $95.7 \pm 0.3$  &    $95.1 \pm 0.4$  &    $94.1 \pm 0.4$  &   $91.3 \pm 0.7$     \\ 
   \hline
\rowcolor{lightgray}   
   \multicolumn{1}{c}{\multirow{2}{*}{\cellcolor{white}\textbf{Drebin NN}}} & \textbf{FPR=0.1} &    $54.9 \pm 13.5$    &  $43.3 \pm 13.5$     &   $29.2 \pm 11.1$   &   $14.7 \pm 6.4$  &  $6.2 \pm 2.2$   &  $7.9 \pm 8.8$      \\ 
 
   \multicolumn{1}{c}{}  & \textbf{FPR=1}   &     $90.3 \pm 1.0$     &    $88.3 \pm 1.1$     &  $85.7 \pm 1.9$  &    $77.7 \pm 3.8$  &    $56.6 \pm 14.8$  &   $46.7 \pm 17.1$     \\ 
   \hline
\rowcolor{lightgray}   
       \multicolumn{1}{c}{\multirow{2}{*}{\cellcolor{white}\textbf{MaMaDroid RF}}} & \textbf{FPR=0.1} &    $60.0 \pm 9.2$    &  $56.4 \pm 8.5$     &   $56.0 \pm 7.1$   &   $53.0 \pm 9.1$  &  $50.1 \pm 9.9$   &  $40.8 \pm 12.0$      \\ 

    \multicolumn{1}{c}{}  & \textbf{FPR=1}    &     $85.8 \pm 1.2$     &    $85.0 \pm 1.2$     &  $84.9 \pm 1.1$  &    $84.7 \pm 1.2$  &    $83.5 \pm 1.3$  &   $80.5 \pm 2.0$     \\ 
   \hline
\rowcolor{lightgray}   
    \multicolumn{1}{c}{\multirow{2}{*}{\cellcolor{white}\textbf{MalScan RF}}} & \textbf{FPR=0.1} &    $67.8 \pm 3.4$    &  $66.0 \pm 10.0$     &   $66.4 \pm 7.5$   &   $67.0 \pm 5.6$  &  $62.8 \pm 6.8$   &  $56.2 \pm 8.4$      \\  

    \multicolumn{1}{c}{}  & \textbf{FPR=1}   &     $88.4 \pm 1.7$     &    $88.5 \pm 1.2$     &  $88.2 \pm 1.3$  &    $87.6 \pm 1.5$  &    $86.4 \pm 1.8$  &   $81.6 \pm 2.2$     \\ 
   \specialrule{1.2pt}{0pt}{0pt}
  \end{tabular}
  }
  \end{table*}

\subsection{Results of Integrity Attacks}
\label{sec:targeted}
We conduct the integrity label spoofing attack with the large-scale dataset \largedata, showing that we can actively misclassify specific benign APKs. As for the DoS large-scale experiments, we emulate the attack to avoid the unethical implications that such activity would cause. In particular, we target the TikTok~\cite{TikTok} APK.
The poisoning variants are generated within the feature space. Specifically, for Drebin features, each variant of the TikTok APK is obtained by introducing one distinct additional feature. For MaMaDroid and MalScan features, each poisoning variant is constructed by perturbing a unique transition probability by an increment of 0.1.
 After the train-test-split process, we generate $q=5, 10, 50, 100, 200, 500, 1000$ unique variants of such benign APK, all labeled as malware.
The attack results using Drebin, MaMaDroid, and MalScan features are shown in \autoref{Table:40K_target_3models} and \autoref{Table:40K_target_2models_mamadroid_malscan}.
For Drebin features, compared to the poison-free baseline, TPR fluctuates by only 1\% (1\% FPR) and 5\% (0.1\% FPR) as the number of poisoning variants increases for both LSVM and GBT. Besides, the TPR of NN also fluctuates by only 1\% and 7\% when FPR is 1\% and 0.1\% respectively, compared to the baseline.
Therefore, the targeted label spoofing attack has a very limited impact on the classification performance of the three classifiers on the whole test set.   
Among the three classifiers, GBT is still the most robust model for the targeted label spoofing attack, with almost no fluctuation of TPR when FPR is 1\% and the number of poisoning variants increases.
In terms of the classification accuracy over the targeted goodware, we find that only 5 poisoning variants (0.015\% of the training set) are sufficient to misclassify TikTok as malware for all three classifiers at 1\% FPR. 
At the low FPR level (FPR=0.1\%), the integrity attack shows more impacts over GBT than LSVM and NN. 
With 10 and 50 poisoning variants (0.03\% and 0.15\% of the training set), ASR on GBT can reach 50\%, significantly higher than LSVM and NN. 
One potential explanation is that the targeted attack is in nature a two-task learning problem, i.e., normal classification over non-targeted APKs and misclassification over the targeted APKs simultaneously. 
Since GBT has a better model capacity than LSVM and NN to describe the adaptive and piecewise non-linear decision boundaries (GBT is more accurate with the FPR level), it hence better learns the two learning tasks, triggering higher TPR and ASR at the same time. 
Since NN has complex and continuous non-linear decision boundaries, it is sensitive to each poisoning variant. As different poisoning variants shift the decision regions differently, ASR of NN fluctuates.
Regarding MaMaDroid, TPR of RF varies by 2\% (1\% FPR) and 10\% (0.1\% FPR) when the number of poisoning variants increases.
Regarding MalScan, TPR of RF varies by 1\% (1\% FPR) and 5\% (0.1\% FPR) when the number of poisoning variants increases. 
When FPR is 1\%, we find 5 poisoning variants induce 100\% of ASR over the targeted TikTok APK for both MaMaDroid and MalScan features. 
Similarly, with FPR = 0.1\%, only 5 poisoning variants can cause 90\% and 80\% of ASR for MaMaDroid and MalScan respectively.
The results confirm the effectiveness of the integrity attack of AndroVenom over the targeted APK.
In conclusion, the integrity attack of AndroVenom can misclassify the targeted goodware as malware with a very tiny ratio of poisoning variants yet keep an accurate classification over the rest of the test samples. 
 
\mypar{Final Remarks on Integrity Attacks.} Given the results, we are able to answer \textbf{RQ4}.
Regardless of the architecture of the classifiers, AndroVenom can make the targeted goodware misclassified as malware by ML-based malware classification models, while maintaining accurate classification on other testing samples.
This observation unveils realistic unethical implications: \emph{Only 5 variants of the benign APK produces excessive false alarms of ML-based malware classification}.

  % Table:40K_target_3models
  \begin{table*}[t]
    \centering 
    \footnotesize
    \definecolor{lightgray}{gray}{0.92}
    \caption{Mean TPR with standard deviation of models trained with Drebin features on \largedata, and Attack Success Rate (ASR) of the integrity attack targeting the TikTok, with an increasing number of poisoning samples.
    We report performances at $0.1\%$ and $1\%$ FPR. All values, except poisoning variants, are in percentage.}
    \label{Table:40K_target_3models}
    \resizebox{0.9\linewidth}{!}{% 
    \begin{tabular}{ccccccccccc}    
     %\cline{4-11} 
     \specialrule{1.2pt}{0pt}{0pt}
     & & & \multicolumn{8}{c}{\textbf{Poisoning Variants}} \\
     %\cline{4-11} 
     &   &  & \textbf{0}      &   \textbf{5}      & \textbf{10}   & \textbf{50}  & \textbf{100}  & \textbf{200} & \textbf{500} & \textbf{1000}      \\
     \hline
     \rowcolor{lightgray}
     \multicolumn{1}{c}{\multirow{2}{*}{\cellcolor{white}\textbf{LSVM}}} & \multicolumn{1}{c}{\multirow{2}{*}{\cellcolor{white}\textbf{FPR=0.1}}} & \multicolumn{1}{c}{\textbf{TPR}} &    $57.0 \pm 6.6$    &  $56.9 \pm 6.5$     &   $56.7 \pm 7.4$   &   $56.3 \pm 7.8$  &  $61.7 \pm 4.9$   &  $61.0 \pm 5.0$  &  $57.3 \pm 5.3$  &  $57.4 \pm 5.4$      \\ 
     %\cline{3-11} 
     \multicolumn{1}{c}{} & \multicolumn{1}{c}{} & \multicolumn{1}{c}{\textbf{ASR}} &    0    &    0   &   0   &  0   &  0   &  0  &  0  &  0      \\
     \cline{2-11}
     \rowcolor{lightgray}
     \multicolumn{1}{c}{\cellcolor{white}} & \multicolumn{1}{c}{\multirow{2}{*}{\cellcolor{white}\textbf{FPR=1}}}  &  \multicolumn{1}{c}{\textbf{TPR}}  &     $92.5 \pm 1.1$     &    $92.1 \pm 1.5$     &  $92.4 \pm 0.8$  &    $91.8 \pm 1.1$  &    $92.2 \pm 1.2$  &   $92.9 \pm 0.8$   &  $92.2 \pm 1.1$    &  $92.0 \pm 1.5$  \\
     %\cline{3-11} 
     \multicolumn{1}{c}{} & \multicolumn{1}{c}{} & \multicolumn{1}{c}{\textbf{ASR}} &    0    &    100   &   100   &  100   &  100   &  100  &  100  &   100     \\ 
     \hline
     \rowcolor{lightgray}
     \multicolumn{1}{c}{\multirow{2}{*}{\cellcolor{white}\textbf{GBT}}} & \multicolumn{1}{c}{\multirow{2}{*}{\cellcolor{white}\textbf{FPR=0.1}}} & \multicolumn{1}{c}{\textbf{TPR}} &    $88.2 \pm 2.2$    &  $83.4 \pm 4.3$     &   $87.0 \pm 3.6$   &   $85.0 \pm 4.2$  &  $86.8 \pm 4.6$   &  $87.9 \pm 3.6$  &  $86.1 \pm 4.8$  &  $85.8 \pm 4.4$      \\ 
     %\cline{3-11} 
     \multicolumn{1}{c}{} & \multicolumn{1}{c}{} & \multicolumn{1}{c}{\textbf{ASR}} &    0    &   0    &   30   &   50  &  80   &  90  &  100  &    100    \\
     \cline{2-11}
     \rowcolor{lightgray}
     \multicolumn{1}{c}{\cellcolor{white}} & \multicolumn{1}{c}{\multirow{2}{*}{\cellcolor{white}\textbf{FPR=1}}}  &  \multicolumn{1}{c}{\textbf{TPR}}  &     $96.6 \pm 0.4$     &    $96.5 \pm 0.2$     &  $96.6 \pm 0.4$  &    $96.5 \pm 0.2$  &    $96.4 \pm 0.5$  &   $96.5 \pm 0.2$   &  $96.6 \pm 0.3$    &  $96.4 \pm 0.4$  \\
     %\cline{3-11} 
     \multicolumn{1}{c}{} & \multicolumn{1}{c}{} & \multicolumn{1}{c}{\textbf{ASR}} &    0    &    100   &   100   &  100   &   100  &  100  &  100  &    100    \\ 
     \hline
     \rowcolor{lightgray}
     \multicolumn{1}{c}{\multirow{2}{*}{\cellcolor{white}\textbf{NN}}} & \multicolumn{1}{c}{\multirow{2}{*}{\cellcolor{white}\textbf{FPR=0.1}}} & \multicolumn{1}{c}{\textbf{TPR}} &    $51.9 \pm 14.7$    &  $54.2 \pm 13.2$     &   $59.6 \pm 8.2$   &   $44.4 \pm 16.2$  &  $54.1 \pm 10.4$   &  $52.9 \pm 11.1$  &  $49.3 \pm 14.0$  &  $54.2 \pm 10.9$      \\ 
     %\cline{3-11} 
     \multicolumn{1}{c}{} & \multicolumn{1}{c}{} & \multicolumn{1}{c}{\textbf{ASR}} &    0    &    60   &   40   &  20   &   30  &  50  &  60  &   40     \\
     \cline{2-11}
     \rowcolor{lightgray}
     \multicolumn{1}{c}{\cellcolor{white}} & \multicolumn{1}{c}{\multirow{2}{*}{\cellcolor{white}\textbf{FPR=1}}}  &  \multicolumn{1}{c}{\textbf{TPR}}  &     $90.4 \pm 1.6$     &    $89.9 \pm 1.2$     &  $90.8 \pm 0.8$  &    $90.1 \pm 1.7$  &    $91.4 \pm 1.0$  &   $90.4 \pm 0.9$   &  $90.1 \pm 1.2$    &  $90.7 \pm 1.1$  \\
     %\cline{3-11} 
     \multicolumn{1}{c}{} & \multicolumn{1}{c}{} & \multicolumn{1}{c}{\textbf{ASR}} &    0    &    80   &   70   &  90   &   70  &  100  &  70  &   80     \\ 
    \specialrule{1.2pt}{0pt}{0pt}
    \end{tabular}
    }
    \end{table*}

  \begin{table}[t]
    \centering
    \footnotesize
    \definecolor{lightgray}{gray}{0.92}
    \caption{Mean TPR with standard deviation of RF trained with MaMaDroid and MalScan features on \largedata, and Attack Success Rate (ASR) of the integrity attack targeting the TikTok, with an increasing number of poisoning samples.
    We report performances at $0.1\%$ and $1\%$ FPR. All values, except poisoning variants, are in percentage.}
    \label{Table:40K_target_2models_mamadroid_malscan}
    \resizebox{\linewidth}{!}{ 
    \begin{tabular}{cccccccc}    
     %\cline{4-8}
     \specialrule{1.2pt}{0pt}{0pt}
     & & & \multicolumn{5}{c}{\textbf{Poisoning Variants}} \\
     %\cline{4-8} 
     &  &  & \textbf{0}      &   \textbf{5}      & \textbf{10}   & \textbf{50}  & \textbf{100}        \\
     \hline
     \rowcolor{lightgray}
      \multicolumn{1}{c}{\multirow{2}{*}{\cellcolor{white}\textbf{MaMaDroid}}} & \multicolumn{1}{c}{\multirow{2}{*}{\cellcolor{white}\textbf{FPR=0.1}}} & \multicolumn{1}{c}{\textbf{TPR}} &    $58.9 \pm 6.6$    &  $55.6 \pm 13.4$     &   $52.2 \pm 9.8$   &   $59.4 \pm 10.4$  &  $55.5 \pm 7.0$         \\ 
     %\cline{3-11} 
      \multicolumn{1}{c}{} & \multicolumn{1}{c}{} & \multicolumn{1}{c}{\textbf{ASR}} &    0    &    90   &   100   &  100   &   100       \\
     \cline{2-8}
     \rowcolor{lightgray}
      \multicolumn{1}{c}{\cellcolor{white}} & \multicolumn{1}{c}{\multirow{2}{*}{\cellcolor{white}\textbf{FPR=1}}}  &  \multicolumn{1}{c}{\textbf{TPR}}  &     $85.4 \pm 1.6$     &    $86.0 \pm 0.9$     &  $85.6 \pm 1.4$  &    $85.9 \pm 0.9$  &    $84.6 \pm 1.4$    \\
     %\cline{3-11} 
       \multicolumn{1}{c}{} & \multicolumn{1}{c}{} & \multicolumn{1}{c}{\textbf{ASR}} &   0   &    100   &   100   &  100   &    100     \\ 
     \hline
     \rowcolor{lightgray}
     \multicolumn{1}{c}{\multirow{2}{*}{\cellcolor{white}\textbf{MalScan}}} & \multicolumn{1}{c}{\multirow{2}{*}{\cellcolor{white}\textbf{FPR=0.1}}} & \multicolumn{1}{c}{\textbf{TPR}} &    $66.8 \pm 3.6$    &  $64.8 \pm 5.3$     &   $64.0 \pm 11.7$   &   $68.0 \pm 2.2$  &  $62.6 \pm 5.5$        \\ 
     %\cline{3-11} 
     \multicolumn{1}{c}{} & \multicolumn{1}{c}{} & \multicolumn{1}{c}{\textbf{ASR}} &   0    &   80    &   100   &  100   &   100    \\
     \cline{2-8}
     \rowcolor{lightgray}
     \multicolumn{1}{c}{\cellcolor{white}} & \multicolumn{1}{c}{\multirow{2}{*}{\cellcolor{white}\textbf{FPR=1}}}  &  \multicolumn{1}{c}{\textbf{TPR}}  &     $88.6 \pm 0.9$     &    $88.2 \pm 1.7$     &  $88.5 \pm 1.9$  &    $89.4 \pm 1.0$  &    $88.4 \pm 1.2$    \\
     %\cline{3-11} 
     \multicolumn{1}{c}{} & \multicolumn{1}{c}{} & \multicolumn{1}{c}{\textbf{ASR}} &   0     &   100    &    100  &   100  &    100      \\ 
    \specialrule{1.2pt}{0pt}{0pt}
    \end{tabular}
    }
    \end{table}

\subsection{Defending against AndroVenom}\label{sec:countermeasure}

\begin{figure}[t]
  \centering
      \includegraphics[width=\linewidth]{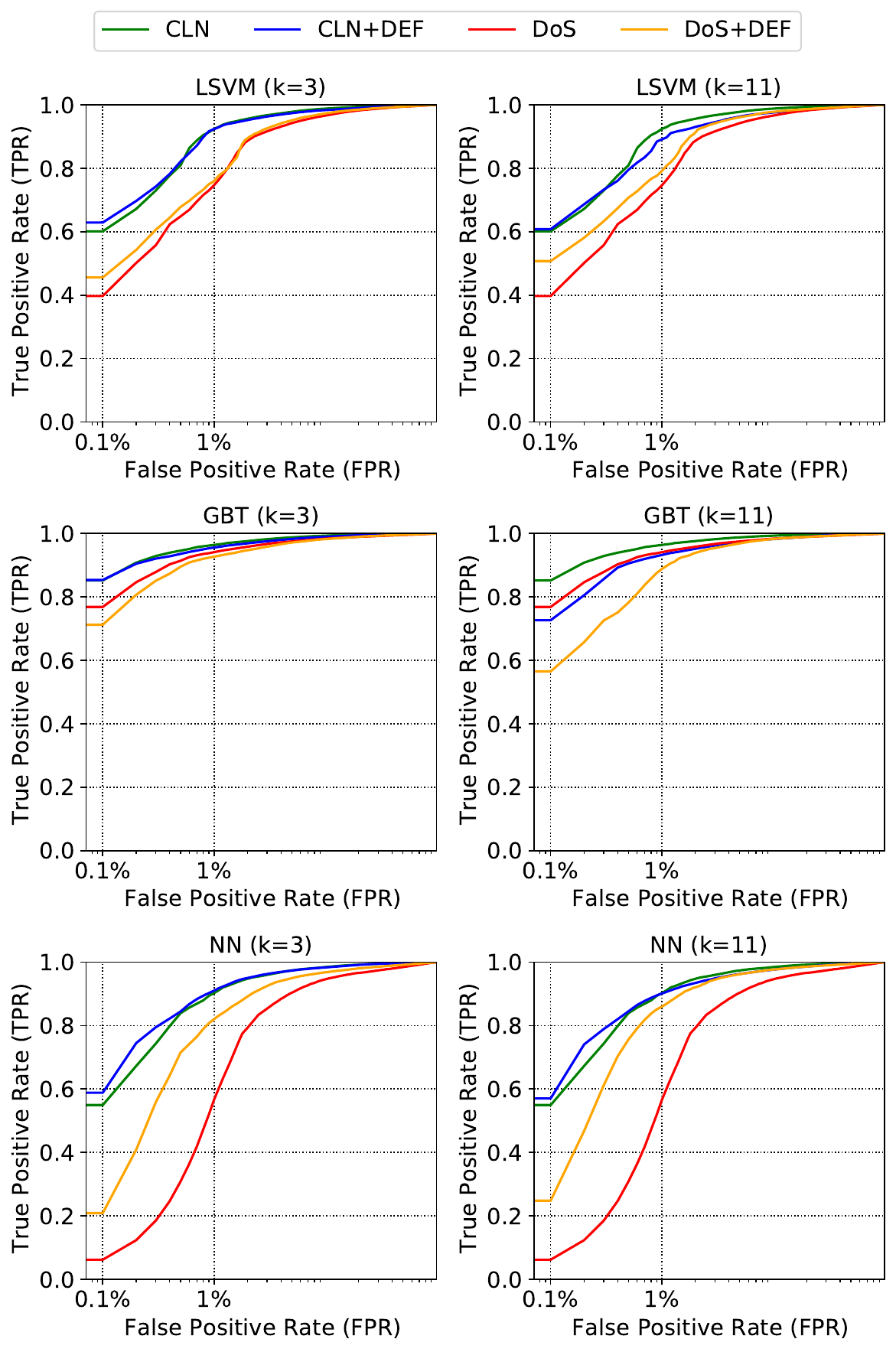}
      \caption{ROC curves of LSVM, GBT, NN using Drebin features with/without the Deep KNN defense strategy (\textbf{DEF}) on \largedata in the attack-free (\textbf{CLN}) and DoS attack (\textbf{DoS}) scenarios. The number of nearest neighbors $k$ is 3 and 11. The poisoning ratio of the DoS attack $p$ is 10\%.}
   \label{fig:ROC_defense_knn_3models}
\end{figure}
%\vspace{-4pt}

\begin{figure}[t]
  \centering
      \includegraphics[width=\linewidth]{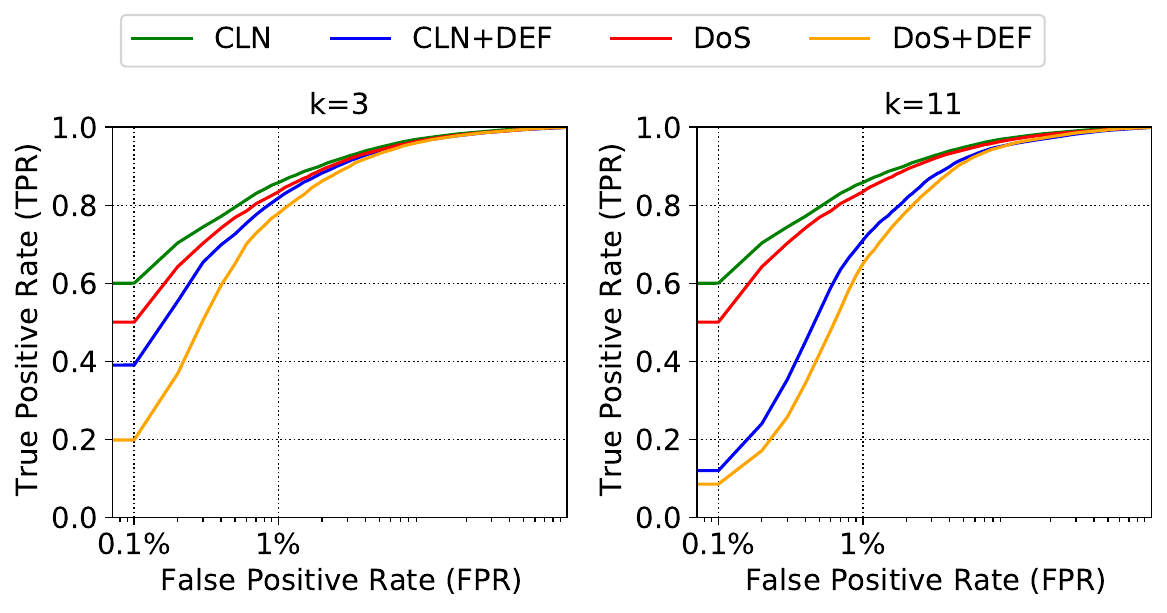}
      \caption{ROC curves of RF using MaMaDroid features with/without the Deep KNN defense strategy (\textbf{DEF}) on \largedata in the attack-free (\textbf{CLN}) and DoS attack (\textbf{DoS}) scenarios. The numbers of nearest neighbors $k$ are 3 and 11. The poisoning ratio of the DoS attack $p$ is 10\%.}
  \label{fig:ROC_defense_knn_RF_mamadroid}
\end{figure}
%\vspace{-4pt}

\begin{figure}[t]
  \centering
      \includegraphics[width=\linewidth]{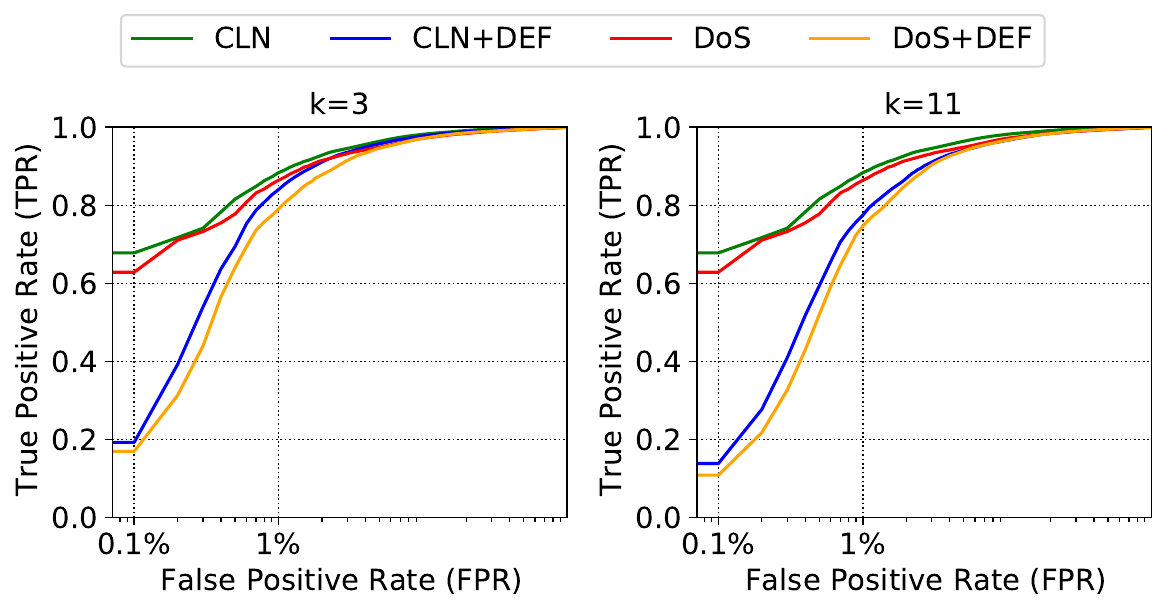}
      \caption{ROC curves of RF using MalScan features with/without the Deep KNN defense strategy (\textbf{DEF}) on \largedata in the attack-free (\textbf{CLN}) and DoS attack (\textbf{DoS}) scenarios. The numbers of nearest neighbors $k$ are 3 and 11. The poisoning ratio of the DoS attack $p$ is 10\%.}
  \label{fig:ROC_defense_knn_RF_malscan}
\end{figure}

We illustrate the defensive effect of the Deep KNN (described in \autoref{subsec:exp_setup}) defense against AndroVenom on \largedata. 
We choose the number of nearest neighbors $k$ as 3 and 11, which were empirically the best hyper-parameters we checked.
We systematically varied the value of $k$ from 1 to 15 in the Deep KNN-based classification method and evaluated the corresponding ASR on the test samples of the target APK. The results showed that ASR remains relatively stable for $k \ge 3$, with the lowest values observed at $k = 3$ and $k = 11$.
We report the defense results against the DoS attack on Drebin features in \autoref{fig:ROC_defense_knn_3models}.
Also, we present the performance of DoS attacks against MaMaDroid and MalScan in presence of the defense in \autoref{fig:ROC_defense_knn_RF_mamadroid} and \autoref{fig:ROC_defense_knn_RF_malscan} respectively. 
\autoref{Table:Defense_target_3models} shows the defense results against the integrity attack on models trained on Drebin, MaMaDroid and MalScan features.
For the DoS attack, we randomly select a percentage $p=10\%$ of benign APKs in the training set, which are used to execute the label spoofing attack. 
For the integrity attack, we fix the number of poisoning variants $q$ as 100. 
In the plots, we show ROC curves of the victim classifier compromised by the DoS attack with/without the Deep KNN defense strategy, noted as DoS+DEF and DoS respectively. 
In addition, we set up a control group to evaluate the impact of the Deep KNN defense strategy over the utility of the poison-free model. 
They include ROC curves of the poison-free classifier with/without the KNN defense, noted as CLN+DEF and CLN respectively. 
Regarding the defense results against the integrity attack, we provide TPR values of the victim classifiers with/without the defense, by freezing FPR = 0.1\% and 1\% respectively. 
The observation for ML models trained on Drebin features can be summarized in two perspectives. 
First of all, we find that Deep KNN does not provide consistent defense across different models, by working only on LSVM and NN models. 
However, GBT models seem to suffer from the inclusion of the defense, experiencing a deterioration of classification accuracy in the DoS attack scenario. 
Specifically, when FPR is 0.1\% and $k$ is 11, TPR of NN increases 20\% but TPR of GBT drops by 20\%.
One possible reason could be that the filtering of training data by Deep KNN also causes a drift inside the training data distribution, which is not ideal for tree-like models like GBT.
Second, we also observe that Deep KNN can not prevent integrity attacks of AndroVenom.
As demonstrated in \autoref{Table:Defense_target_3models}, there is no defensive effect on LSVM and the ASR of NN even increases with the defense. 
Although the ASR of GBT decreases when FPR is 0.1\%, the defense has no effect when FPR is 1\%. 
As shown by \autoref{fig:ROC_defense_knn_RF_mamadroid} and \autoref{fig:ROC_defense_knn_RF_malscan}, the Deep KNN defense causes significant TPR drop on both the poison-free and the attacked RF model in the DoS attack mode. In the integrity attack mode, with FPR = 0.1\% and 100 poisoning variants, the Deep KNN defense can reduce ASR. 
However, it also causes the TPR to drop significantly.
When FPR = 1\%, it does not show mitigation effects. 
This observation denotes that the introduced defense method brings extra harms to the utility of the ML model, yet without offering a strong protection. 
A possible reason behind this suboptimal result is due to the proximity in the feature space of poisoned samples and regular ones.
The poisoned samples are all close to each other in the feature space, being consistently labeled as malicious.
Consequently, most of its nearest samples still carry the attack-desired malicious label, which circumvents the defensive mechanism using Deep KNN.

\begin{table*}[t]
    \scriptsize
    \centering
    \definecolor{lightgray}{gray}{0.92}
    \caption{Mean TPR with standard deviation of models trained with Drebin, MaMaDroid, and MalScan features on \largedata, and Attack Success Rate (ASR) of the integrity attack targeting the TikTok, with/without the Deep KNN defense strategy.
    We report performances at $0.1\%$ and $1\%$ FPR. All values are in percentage except the number of poisoning variants ($q$) and nearest neighbors ($k$).}
    \label{Table:Defense_target_3models}
    \resizebox{0.8\linewidth}{!}{ 
    \begin{tabular}{cccccc|ccc}    
     % \cline{4-9}
     \specialrule{1.2pt}{0pt}{0pt}
     & & & \multicolumn{3}{c|}{\textbf{q=0}} & \multicolumn{3}{c}{\textbf{q=100}} \\
     %\cline{4-9} 
     &   &  & \textbf{no defense}      &   \textbf{k=3}      & \textbf{k=11}   & \textbf{no defense}  & \textbf{k=3}  & \textbf{k=11}       \\
     \hline
     \rowcolor{lightgray}
     \multicolumn{1}{c}{\multirow{2}{*}{\cellcolor{white}\textbf{Drebin LSVM}}} & \multicolumn{1}{c}{\multirow{2}{*}{\cellcolor{white}\textbf{FPR=0.1}}} & \multicolumn{1}{c}{\textbf{TPR}} &    $57.0 \pm 6.6$    &  $62.9 \pm 6.9$     &   $60.8 \pm 7.7$    &  $61.7 \pm 4.9$   &  $61.3 \pm 4.9$  &  $64.1 \pm 3.8$   \\ 
     %\cline{3-11} 
     \multicolumn{1}{c}{} & \multicolumn{1}{c}{} & \multicolumn{1}{c}{\textbf{ASR}} &    0    &   0    &   0    &  0   &  0  &   0      \\
     \cline{2-9}
     \rowcolor{lightgray}
     \multicolumn{1}{c}{\cellcolor{white}} & \multicolumn{1}{c}{\multirow{2}{*}{\cellcolor{white}\textbf{FPR=1}}}  &  \multicolumn{1}{c}{\textbf{TPR}}  &     $92.5 \pm 1.1$     &    $92.3 \pm 1.0$     &  $89.1 \pm 3.1$    &    $92.2 \pm 1.2$  &   $91.7 \pm 0.8$   &  $90.3 \pm 1.4$     \\
     %\cline{3-11} 
     \multicolumn{1}{c}{} & \multicolumn{1}{c}{} & \multicolumn{1}{c}{\textbf{ASR}} &    0    &   0    &    0    &  100   &  100  &   90      \\ 
     \hline
     \rowcolor{lightgray}
     \multicolumn{1}{c}{\multirow{2}{*}{\cellcolor{white}\textbf{Drebin GBT}}} & \multicolumn{1}{c}{\multirow{2}{*}{\cellcolor{white}\textbf{FPR=0.1}}} & \multicolumn{1}{c}{\textbf{TPR}} &    $88.2 \pm 2.2$    &  $85.3 \pm 4.7$     &   $72.6 \pm 7.9$     &  $86.8 \pm 4.6$   &  $78.4 \pm 7.5$  &  $69.3 \pm 10.2$       \\ 
     %\cline{3-11} 
     \multicolumn{1}{c}{} & \multicolumn{1}{c}{} & \multicolumn{1}{c}{\textbf{ASR}} &    0    &   0    &    0    &  80   &  30  &  0     \\
     \cline{2-9}
     \rowcolor{lightgray}
     \multicolumn{1}{c}{\cellcolor{white}} & \multicolumn{1}{c}{\multirow{2}{*}{\cellcolor{white}\textbf{FPR=1}}}  &  \multicolumn{1}{c}{\textbf{TPR}}  &     $96.6 \pm 0.4$     &    $95.6 \pm 0.4$     &  $93.3 \pm 0.5$    &    $96.4 \pm 0.5$  &   $95.3 \pm 0.4$   &  $92.8 \pm 0.8$      \\
     %\cline{3-11} 
     \multicolumn{1}{c}{} & \multicolumn{1}{c}{} & \multicolumn{1}{c}{\textbf{ASR}} &    0    &   0    &    0     &   100  &   100  &  100      \\ 
     \hline
     \rowcolor{lightgray}
     \multicolumn{1}{c}{\multirow{2}{*}{\cellcolor{white}\textbf{Drebin NN}}} & \multicolumn{1}{c}{\multirow{2}{*}{\cellcolor{white}\textbf{FPR=0.1}}} & \multicolumn{1}{c}{\textbf{TPR}} &    $51.9 \pm 14.7$    &  $58.8 \pm 12.9$     &   $57.0 \pm 15.0$     &  $54.1 \pm 10.4$   &  $54.1 \pm 11.7$  &  $57.9 \pm 10.0$       \\ 
     %\cline{3-11} 
     \multicolumn{1}{c}{} & \multicolumn{1}{c}{} & \multicolumn{1}{c}{\textbf{ASR}} &    0    &   0    &    0     &   30  &  60  &  80      \\
     \cline{2-9}
     \rowcolor{lightgray}
     \multicolumn{1}{c}{\cellcolor{white}} & \multicolumn{1}{c}{\multirow{2}{*}{\cellcolor{white}\textbf{FPR=1}}}  &  \multicolumn{1}{c}{\textbf{TPR}}  &     $90.4 \pm 1.6$     &    $91.0 \pm 1.0$     &  $90.1 \pm 1.0$    &    $91.4 \pm 1.0$  &   $91.1 \pm 1.2$   &  $90.5 \pm 0.7$     \\
     %\cline{3-11} 
     \multicolumn{1}{c}{} & \multicolumn{1}{c}{} & \multicolumn{1}{c}{\textbf{ASR}} &    0    &   0    &    0    &   70  &   100  &   100    \\ 
     \hline
     \rowcolor{lightgray}
     \multicolumn{1}{c}{\multirow{2}{*}{\cellcolor{white}\textbf{MaMaDroid RF}}} & \multicolumn{1}{c}{\multirow{2}{*}{\cellcolor{white}\textbf{FPR=0.1}}} & \multicolumn{1}{c}{\textbf{TPR}} &    $58.9 \pm 6.6$    &  $35.3 \pm 10.6$     &   $12.1 \pm 3.7$    &  $55.5 \pm 7.0$   &  $30.1 \pm 10.6$  &  $9.2 \pm 1.2$   \\ 
   %\cline{3-11} 
     \multicolumn{1}{c}{} & \multicolumn{1}{c}{} & \multicolumn{1}{c}{\textbf{ASR}} &    0    &   0    &   0    &   100   &  70  &    0     \\
   \cline{2-9}
   \rowcolor{lightgray}
      \multicolumn{1}{c}{\cellcolor{white}} & \multicolumn{1}{c}{\multirow{2}{*}{\cellcolor{white}\textbf{FPR=1}}}  &  \multicolumn{1}{c}{\textbf{TPR}}  &   $85.4 \pm 1.6$     &    $81.1 \pm 1.9$     &  $70.1 \pm 2.9$    &    $84.6 \pm 1.4$  &   $80.7 \pm 2.1$   &  $67.9 \pm 4.3$     \\
   %\cline{3-11} 
     \multicolumn{1}{c}{} & \multicolumn{1}{c}{} & \multicolumn{1}{c}{\textbf{ASR}} &   0    &   0    &     0   &  100   &  100  &    100     \\ 
   \hline
   \rowcolor{lightgray}
    \multicolumn{1}{c}{\multirow{2}{*}{\cellcolor{white}\textbf{MalScan RF}}} & \multicolumn{1}{c}{\multirow{2}{*}{\cellcolor{white}\textbf{FPR=0.1}}} & \multicolumn{1}{c}{\textbf{TPR}} &    $66.8 \pm 3.6$    &  $18.5 \pm 3.6$     &   $14.3 \pm 3.4$    &  $62.6 \pm 5.5$   &  $15.2 \pm 3.0$  &  $12.7 \pm 2.2$   \\ 
   %\cline{3-11} 
     \multicolumn{1}{c}{} & \multicolumn{1}{c}{} & \multicolumn{1}{c}{\textbf{ASR}} &   0     &   0    &   0    &   100   &  0  &  0       \\
   \cline{2-9}
   \rowcolor{lightgray}
      \multicolumn{1}{c}{\cellcolor{white}} & \multicolumn{1}{c}{\multirow{2}{*}{\cellcolor{white}\textbf{FPR=1}}}  &  \multicolumn{1}{c}{\textbf{TPR}}  &   $88.6 \pm 0.9$     &    $85.1 \pm 1.4$     &  $78.6 \pm 2.2$    &    $88.4 \pm 1.2$  &   $83.3 \pm 2.5$   &  $77.3 \pm 3.1$     \\
   %\cline{3-11} 
     \multicolumn{1}{c}{} & \multicolumn{1}{c}{} & \multicolumn{1}{c}{\textbf{ASR}} &   0    &     0  &     0   &   100  &  100  &  100       \\ 
   \specialrule{1.2pt}{0pt}{0pt}

    \end{tabular}
    }
    \end{table*}
 
\section{Related Work}
\label{sec:related}

Inspired by Cin{\`a} et al.~\cite{cina2023wild}, we categorize in \autoref{Table:related poisoning attacks} the existing poisoning attacks on malware classification into six dimensions: the attacker's goal, the attacker's knowledge, the attacker's manipulation surface, the attacker's capability, the operating system of the dataset, and the feature used for malware classification.
\emph{(1)} The attacker's goal can be either a DoS attack or an integrity attack, which is presented in \autoref{sec:background}.
\emph{(2)} The attacker's knowledge can be either a white-box attack or a black-box attack. A white-box attack has complete knowledge about the targeted system, including the training data, the test data, and the targeted model. This attack performs a worst-case scenario. A black-box attack assumes that the attacker has no knowledge of the targeted model.
\emph{(3)} The attacker's manipulation surface can be either a feature space manipulation or a problem space manipulation. A feature space manipulation theoretically modifies only the feature space of samples, such as feature vectors. A problem space manipulation makes changes directly on real samples. Hence, the problem space manipulation can test the validity of the feature space manipulation. 
\emph{(4)} The attacker's capability can be either a clean-label attack, a dirty-label attack, or a label-flip attack, which is introduced in \autoref{sec:background}.
\emph{(5)} The main operating systems of the dataset for malware classification are Android and Windows. 
\emph{(6)} Since our work focuses on Android malware detection, we only summarize the Android features used in previous works. Apart from Drebin and MaMaDroid feature set presented in \autoref{sec:background}, some works use only permissions~\cite{Taheri2024Unveiling} or API call graphs~\cite{Xi21Graph} for their feature extractors.

Some works~\cite{Severi21Explanation-Guided,Xi21Graph,Kuppa21Adversarial,Shan22Poison,Li22Backdoor,Yang23Jigsaw,Noppel23Disguising,Tian23Sparsity,Qi23Towards,zhan2025practical} report an integrity attack, which aims to activate misclassification on specific malware. 
Besides, the classifier still contributes normal classification behaviors over the other files. With an explanation AI-based technique, \cite{Severi21Explanation-Guided} can perform better search over important features to poison. 
Instead, \cite{Chen18Automated,Demontis19Why,Aryal22Analysis,Taheri2024Unveiling,McFadden2024ACSAC} involve the DoS attack mode. 
Notably, \cite{McFadden2024ACSAC} is established based on Android malware classification use cases. 
By injecting label-flipping noise into the active learning process, the proposed attack achieves to harm the overall availability of the targeted ML system.  
Most works of Android malware classification choose Drebin or MaMaDroid as the features.
While some previous poisoning attack methods in ML-based malware classifiers concentrate on tampering the feature space, our study focuses on a realistic problem space attack scenario. 
Specifically, we deliver the attack via injecting malicious files to activate misannotation of benign training samples, yet without compromising the utility of the modified apps or the extracted features.

\begin{table} [!htb]
\definecolor{lightgray}{gray}{0.92}
\caption{Poisoning attacks on malware classification categorized by the attacker's goal (\textbf{S} = DoS / \textbf{I} = Integrity), knowledge (\textbf{W} = White-box / \textbf{B} = Black-box), manipulation surface (\textbf{F} = Feature space / \textbf{P} = Problem space), capability (\textbf{CL} = Clean-Label / \textbf{DL} = Dirty-Label / \textbf{LF} = Label-Flip), operating system (\textbf{AD} = Android / \textbf{WD} = Windows), and feature (\textbf{D} = Drebin / \textbf{M} = MaMaDroid / \textbf{G} = API call graph / \textbf{E} = Permission).}
\begin{center}
\resizebox{\linewidth}{!} 
{
\begin{tabular}{c  c c c c c c c}
  \specialrule{1.2pt}{0pt}{0pt}
   \textbf{WORK} & \textbf{YEAR} & \textbf{GOAL} & \textbf{KNOW} & \textbf{MANI} & \textbf{CAPA} & \textbf{OS} & \textbf{FEAT}\\
%    \cline{2-7}
  \hline
  \rowcolor{lightgray}
	  Chen et al. \cite{Chen18Automated} & 2018  & S  & W   &  F  & DL  &  AD  &  D \\ 
	
    Demontis et al. \cite{Demontis19Why} & 2019  &  S &  W  &  F  & DL  &  AD  &  D \\ 
\rowcolor{lightgray}	
	  Severi et al. \cite{Severi21Explanation-Guided} & 2021  &  I &  B  &  P  & CL  &  AD  &  D \\ 
	
	  GTA \cite{Xi21Graph} & 2021  &  I   &  B  &  P  & DL  &  AD  &  G \\ 
	\rowcolor{lightgray}
    Kuppa et al. \cite{Kuppa21Adversarial} & 2021  &  I &  B  &  P  & CL  &  WD  &  - \\
    
    Shan et al. \cite{Shan22Poison} & 2022  &  I &  B  &  F  & CL  &  WD  &  - \\
\rowcolor{lightgray}
    Li et al. \cite{Li22Backdoor} & 2022  &  I &  B  &  P  & DL  &  AD &  D + M \\

    Aryal et al. \cite{Aryal22Analysis} & 2022  & S  & B   & -   & LF  &  WD &  - \\
\rowcolor{lightgray}
    Jigsaw Puzzle \cite{Yang23Jigsaw} & 2023  &  I &  B  &  P  & CL  &  AD  &  D \\

    Noppel et al. \cite{Noppel23Disguising} & 2023  &  I &  W  &  F  & DL  &  AD  &  D \\
\rowcolor{lightgray}
    Tian et al. \cite{Tian23Sparsity} & 2023  &  I &  W + B  &  P  & CL  &  AD  &  D \\

    Qi et al. \cite{Qi23Towards} & 2023  &  I &  B  &  F  & CL  &   WD   &  - \\
\rowcolor{lightgray}
    Taheri et al. \cite{Taheri2024Unveiling} & 2024  & S & W & F & DL & AD & E \\
    
    McFadden et al. \cite{McFadden2024ACSAC} & 2024  & S & B & - & LF & AD & D \\
\rowcolor{lightgray}
    Zhan et al. \cite{zhan2025practical} & 2025  & I & B & P  & CL & WD & - \\
\specialrule{1.2pt}{0pt}{0pt}
	 \end{tabular}
	 \label{Table:related poisoning attacks}
}
\end{center}
% \vspace{-4mm}
\end{table}

\section{Conclusions}
\label{sec:takeaways}

In this paper, we introduced potential and highly realistic threats originating from the blind trust afforded to Antivirus (AV) engines, which enables dangerous poisoning attacks against Machine Learning (ML) models. We presented AndroVenom, a framework for constructing samples that are consistently labeled as malicious while remaining entirely devoid of harmful functionality. By doing so, we have demonstrated a realistic poisoning attack capable of compromising \emph{any} ML model trained on crowd-sourced AV data.

This is achieved by coercing various AV engines into generating false labels through the injection of specific malicious signatures into benign APKs. Such manipulation not only triggers a malicious misclassification but also grants attackers control over the malware family to which these samples are attributed. Our study reveals that these label spoofing attacks can be orchestrated against a wide spectrum of ML models for Android malware classification that rely on static features.

At the time of writing, none of the prominent state-of-the-art ML-based Android malware classification methods include an analysis of the \texttt{res} folder within the APK~\cite{ArpSHGR14,Li22Backdoor,JEON20201,Cheng07MobiSys,ruggia2022,aonzo2017low,onwuzurike2019mamadroid}. This omission makes our injection invisible to all current preprocessing and feature extraction algorithms. Furthermore, the sheer volume of crowd-sourced data makes manual verification of AV-based annotations practically infeasible, presenting a significant challenge for mitigation.

\mypar{Attack Impact and Distribution Risks.}
We evaluated both Denial-of-Service (DoS) and integrity attacks. While we observed a consistent drop in test-time performance across various models, the efficacy of our integrity attacks is particularly noteworthy: only 5 label-spoofed samples are required to force the misclassification of a specific benign APK.
The consequences extend beyond model accuracy. Given the time and cost required to manually verify these false alarms, users may hesitate to download APKs affected by AndroVenom for extended periods. Consequently, the reputation of the targeted APK may suffer, significantly impeding legitimate distribution.

\mypar{Future Improvements and Limitations.}
This work demonstrates the feasibility of label spoofing and raise concerns regarding the trustworthiness of AV-based annotation pipelines. As shown in \autoref{sec:androvenom}, attackers possess significant flexibility regarding the types of malicious files they can inject. This scenario can be made even more stealthy by leveraging diverse malicious patterns to produce multiple file variants. We argue that there is much to explore regarding how these attacks can be further optimized. For instance, while we injected entire files easily recognizable by their hash, future work could identify the minimum indispensable modification required to maintain control over label spoofing and family attribution. Such advanced attack techniques would be highly specific to the targeted operating system and file format.

One limitation of our current attack is the necessity of modifying the APK to inject the malicious file, and this process inevitably invalidates the digital signature. AV companies could potentially mitigate this for a subset of legitimate apps by creating allow-lists based on original digital signatures. Furthermore, even if an attacker modifies an APK without altering its behavior (as with AndroVenom), developers may implement integrity checks within the execution logic to detect tampering. In such cases, the modified app might exhibit different feature profiles during dynamic analysis.

Finally, we consciously chose not to emphasize remediation techniques for the following reasons. Although certain ML techniques can suppress label-flipping noise during training~\cite{Peri2020ECCV, levine2021deep, pmlr-v162-wang22m, pmlr-v162-chen22k, RezaeiBCF23, Xie23CCS}, our core message is to highlight the inherent vulnerability of relying blindly on crowd-sourced AV annotations. These ML-based defenses are typically applied during the training process and do not address the root cause: the submission of modified files to AV-based platforms.

The vulnerability in the AV-based annotation process is difficult to curb in practice, as mitigation requires full support from the entire AV vendor community. Our findings suggest that researchers should enrich the feature sets used for classification. For example, integrating resource files into the feature set could dilute or flag the impact of misannotated samples. Future research will explore further countermeasures, such as clustering or ensemble learning-based defenses~\cite{levine2021deep, Jia2020IntrinsicCR}, to identify suspicious samples that share near-identical static features yet are assigned conflicting labels. While these methods help detect outliers, establishing a certifiable, untamperable annotation process remains the ultimate bottleneck.

\section{Acknowledgement}
The work of the first author was conducted under the supervision of Prof. Farid Nait-Abdesselam and was primarily supported by a Fellowship Grant from Université Paris Cité.
The work was also partially supported by the French National Research Agency (Agence Nationale de la Recherche) via grants ``ANR-22-PECY-0007'' (DefMal) and ``ANR-23-IAS4-0001'' (CKRISP).
%\newpage

\bibliographystyle{unsrt}
% {\footnotesize\bibliography{biblio}}
\bibliography{biblio}

\end{document}